\documentclass[journal,comsoc]{IEEEtran}

\usepackage[T1]{fontenc}

\ifCLASSINFOpdf
\else
\fi
\usepackage{color}
\usepackage{xcolor}
\usepackage{algorithmic}
\usepackage{setspace}
\usepackage{cite}
\usepackage{amssymb}
\usepackage[ruled,linesnumbered]{algorithm2e}

\usepackage{amsmath,amsthm}

\usepackage{stfloats}
\usepackage{cuted}
\usepackage{graphicx}
\usepackage{epsfig}
\usepackage{subfigure}
\usepackage{url}
\usepackage{multirow}
\usepackage{makecell}
\usepackage{enumerate}
\usepackage{cite}
\usepackage{multicol} 
\usepackage{mathtools,cuted,lipsum}
\usepackage[numbers]{natbib}
\graphicspath{{pic/}}
\usepackage[cmintegrals]{newtxmath}

\begin{document}
%
\title{Secure Transmission by Leveraging Multiple Intelligent Reflecting Surfaces in MISO Systems}
%
%
%

\author{Jian Li,~\IEEEmembership{Student Member,~IEEE,}
        Lan Zhang,~
        Kaiping Xue,~\IEEEmembership{Senior Member,~IEEE,}
        Yongdong Zhang,~\IEEEmembership{Senior Member,~IEEE,}
        and~Yuguang Fang,~\IEEEmembership{Fellow,~IEEE}
}

\maketitle

\begin{abstract}
Due to the broadcast nature of wireless communications, physical layer security has always been a fundamental but challenging problem.
Recent advance of Intelligent Reflecting Surface (IRS) introduces a new dimension for secure communications by reconfiguring the transmission environments. In this paper, we devise a secure transmission scheme for multi-user MISO systems by leveraging multiple collaborative IRSs.
Specifically, to guarantee the worst-case achievable secrecy rate among multiple legitimate users, we formulate a max-min problem that can be solved by an alternative optimization method to decouple it into multiple sub-problems. Based on semidefinite relaxation and successive convex approximation, each sub-problem can be further converted into convex problem and easily solved.
Extensive experimental results demonstrate that our proposed scheme can adapt to complex scenarios for multiple users and achieve significant gain in terms of achievable secrecy rate.
The gap between the secrecy rate from our proposed scheme and that from the traditional sum-rate maximization and our results show that the secrecy rate obtained from our scheme converges to that achieved from the traditional sum-rate maximization when increasing the number of elements on IRSs.
\end{abstract}


%
\IEEEpeerreviewmaketitle

\section{Introduction}
Due to the broadcast nature of wireless signals, it is vulnerable for user's confidential messages in wireless communications. To safeguard communication security, physical layer security, which can be traced back to 1970's Wyner's seminal work \cite{chraiti2017achieving,li2013transmit,he2014mimo}, has been regarded as a key complement to higher-layer encryption techniques \cite{wang2019secure,guan2020intelligent}.
In traditional communication systems, beamforming and Artificial Noise (AN) are considered as two effective approaches to defending against wiretapping channel and achieving secure communication \cite{zou2015improving,shiu2011physical,zhu2014joint,zhu2015improving}. By exploiting multiple antennas and shaped beams, beamforming technology can be implemented to direct the signal towards the legitimate user and thus reduce the signal leakage.
In addition to beamforming, AN technology can create significant interference and lower the SINR at eavesdroppers by properly designing AN signals. Thus, the achievable secrecy rate, which is a widely used criterion to capture the difference between mutual information of ``Alice-Bob'' intended channel and ``Alice-Eve'' eavesdropping channel and measure the security level, can be effectively improved especially when the channel state of transmitter-user and transmitter-eavesdropper are highly correlative during the transmission.
Nevertheless, due to the complex environment of wireless communications, the proposed approaches do not always work as expected.

As a promising technology to achieve smart radio environment/intelligent radio environment in next generation cellular systems \cite{di2020smart,di2019smart}, Intelligent Reflecting Surfaces (IRSs) can provide reconfigurable signal propagation environments to support cost-effective and power-efficient wireless communication services.
Specifically, IRS is a metasurface composed of a large number of passive reflecting elements, which consumes much lower energy compared with traditional active relays/transceivers \cite{wu2019towards,wu2019intelligent}. By adaptively adjusting the reflection amplitude and/or phase shift of each element, the strength and direction of the incident electromagnetic wave becomes highly controllable.
Thus, IRS is regarded as a novel solution to achieving configurable wireless transmission environment/intelligent radio environment/wireless 2.0 with low hardware/energy cost, and has been applied in various wireless applications such as coverage extension, interference cancellation, and energy efficiency enhancement \cite{wu2019towards,di2020smart,zhang2020augmenting}.
Particularly, Zhang \textit{et al.} \cite{zhang2019tunable,zhang2020augmenting} articulate how to augment transmission environments with cheap reflectors without damaging aesthetic nature of users' surrounding to boost transmission performance and present high layer implementation consideration.
Due to the aforementioned advantages, the IRS-assisted communication systems have great potential to enhance physical layer security.
By jointly optimizing operations on transmitter and passive reflecting elements of IRS, the transmitter-user channel state can be reconfigured to lower the signal leakage to eavesdroppers. Intuitively, users geographically close to the IRS are more likely beneficial from IRS by receiving the tuned signal, whose achievable secrecy rate can be significantly improved.


Recently, some efforts have been made to study IRS-assisted systems for physical layer security. Cui \textit{et al.} \cite{cui2019secure} investigated an IRS-aided secure wireless communication systems where a simple scenario with one eavesdropper is investigated to show the effectiveness of IRS. To explore the effectiveness of traditional approach in IRS-assisted scenarios, Guan \textit{et al.} \cite{guan2020intelligent} further considered AN in an IRS-assisted system, whose performance was verified with the significant gain on secrecy rate. To improve the algorithm efficiency, Yu \textit{et al.} \cite{yu2019enabling} proposed an efficient algorithm adopting block coordinate descent and minorization maximization method for faster convergence especially for large-scale IRS. Dong \textit{et al.} \cite{dong2020secure} also adopted a similar efficient design for Mutiple-Input Mutiple-Output (MIMO) systems. Lyu \textit{et al.} \cite{lyu2020irs} considered a potential IRS threat called IRS jamming attack, which can leverage signals from a transmitter by controlling reflected signals to diminish the signal-to-interference-plus-noise ratio at the user. Since the IRS jammer operates in a passive way, it can be even harder to defend. Xu \textit{et al.} \cite{xu2019resource} studied resource allocation design in multi-user scenarios and also considered AN at transmitter.
However, the aforementioned efforts only focus on the proof-of-concept study by implementing a single IRS.
Thus, the security gain from leveraging multiple collabortive IRSs has not been explored yet, and it is also paramount to jointly optimize wireless transmission environments and allocate resources for legitimate users in multiple IRSs-assisted systems.

To enhance transmission the security from users, in this paper, we study secure transmission schemes for multi-user Mutiple-Input Singe-Output (MISO) systems assisted by multiple collaborative IRSs. 
The main contributions of this paper are summarized as follows.
\begin{itemize}
  \item To deal with the threat from potential eavesdroppers, we propose a secure communication scheme in multiple IRSs-assisted systems. Considering the security requirement for each legitimate user, we formulate a max-min problem to maximize the lower bound of the secrecy rate to optimize the worst performance of multiple users in case eavesdroppers ``steal'' useful information from a certain user.
  \item To solve the formulated max-min problem, we adopt an alternating algorithm to decouple it into multiple sub-problems. In each iteration, we apply Semi-Definite Relaxation (SDR) and Successive Convex Approximation (SCA) method to solve a convex optimization problem.
  \item To verify the effectiveness of the proposed scheme, extensive numerical evaluations are conducted.  Compared with the traditional IRS scheme with beamforming, 
      the results show that the proposed scheme can achieve significant improvement, and the additional AN can improve achievable secrecy rate especially in multi-user scenarios. Moreover, we compare our scheme with the traditional sum-rate problem to show the gap of security performance. With the increase in the number of elements on IRSs, we show that the performance from our scheme converges to that from the sum-rate maximization in terms of sum of secrecy rate.
\end{itemize}

\textit{Symbol Notation:} Boldface lowercase and uppercase letters denote vectors and matrices, respectively. For a vector $\mathbf{a}$, $||\mathbf{a}||$ denotes the Euclidean norm. For matrix $\mathbf{A}$, the conjugate transpose, rank and trace of $\mathbf{A}$ are denoted as $\mathbf{A}^H$, ${\rm Rank}(\mathbf{A})$ and ${\rm Tr}(\mathbf{A})$, respectively. For a complex number $c$, $|c|$ denotes the modulus. $angle(c)$ denotes the phase of the complex value $c$. The set of $n$-by-$m$ real matrices, complex matrices and complex Hermitian matrices are denoted as $\mathbb{R}^{n\times m}$, $\mathbb{C}^{n\times m}$ and $\mathbb{H}^{n\times m}$, respectively. $\mathbf{A}\succeq0$ means $\mathbf{A}$ is a positive semidefinite matrix, and $\mathcal{N}(\mu, \sum)$ denotes the Gaussian distribution with mean $\mu$ and covariance matrix $\sum$.
\section{System Model}

\begin{figure}[t]
  \centering
  \includegraphics[width=0.93\linewidth]{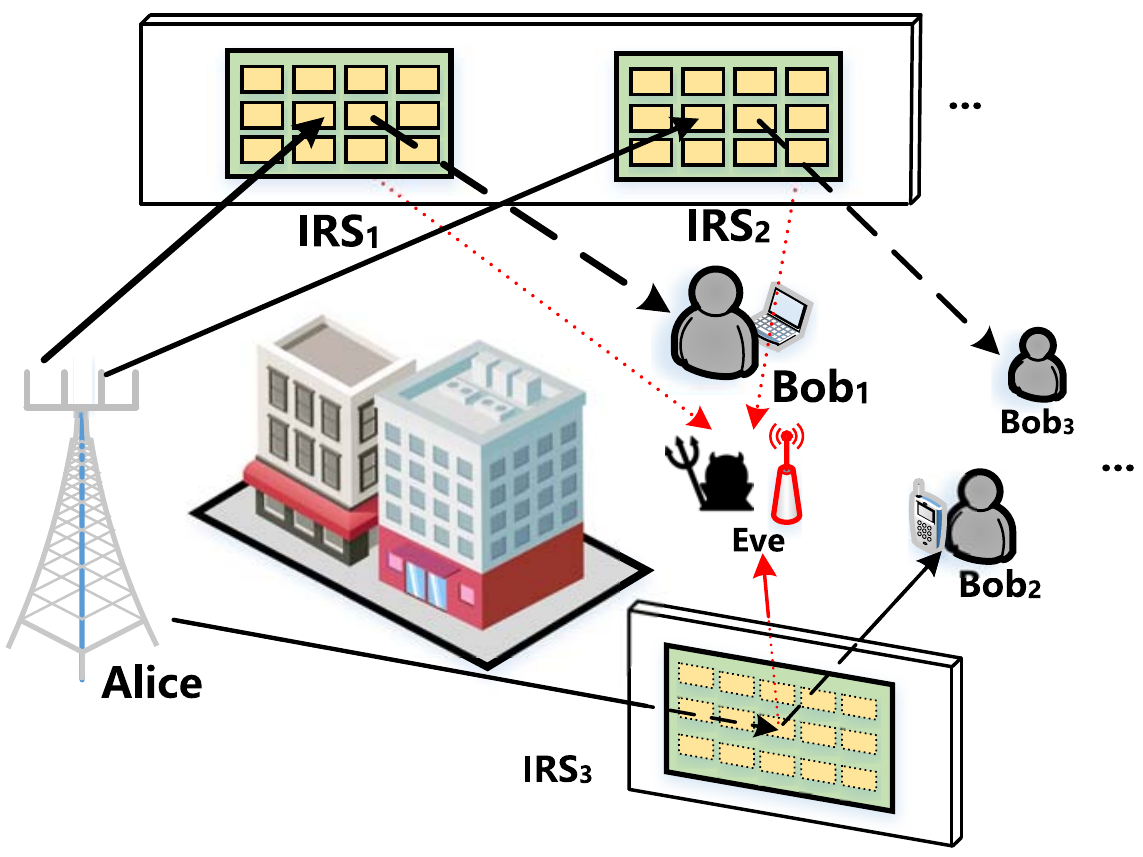}
  \caption{System model of IRSs-assisted secure transmissions.}
  \label{SystemModel}
\end{figure}

We consider a wireless communication system as shown in Fig. \ref{SystemModel}, a base station equipped with $M$ antennas intends to transmit secure messages to $I$ legitimate users equipped with single antenna. Moreover, $K$ IRSs have been deployed in advance to assist wireless communications, and each IRS has $N$ reflecting elements.

\textbf{\textit{Adversary Model:}}
With respect to the transmitted secure messages, one eavesdropper (Eve) wants to wiretap/intercept transmitted signals, and further crack the secure messages to steal users' private information or hack users' equipments. To eliminate the potential threat from the eavesdropper and protect the security of legitimate users, the base station and IRSs need to cooperatively transmit signals to increase received signal power at legitimate users while mitigating the signal leakage at the eavesdropper. In this paper, we attempt to adjust the transmission strategy both at base station and on IRSs to enhance the security level.


\textbf{\textit{Channel Model:}}
There are two parts of a channel experienced from base station to user/Eve, i.e., direct (transmitter-user/Eve) channel and reflecting (transmitter-IRS-user/Eve) channel.
The composite reflecting channel is modeled as a combination of three components, i.e., the base station to IRS link, IRS's reflection with phase shift and IRS to user/Eve link.
The equivalent channels from the base station to the $k$-th IRS, the $i$-th user and Eve are denoted by $\boldsymbol{G}^H_k\in\mathbb{C}^{N\times M}$, $\boldsymbol{h}_i^H\in\mathbb{C}^{1\times M}$, $\boldsymbol{h}^H_e\in\mathbb{C}^{1\times M}$, respectively. The equivalent channels from the $k$-th IRS to the $i$-th user and Eve are denoted by $\boldsymbol{g}^H_{i,k}\in\mathbb{C}^{1\times N}$, $\boldsymbol{g}^H_e\in\mathbb{C}^{1\times N}$, respectively.
Since IRS is a passive reflecting device, we consider a Time Division Duplexing (TDD) protocol for uplink and downlink transmissions and quasi-static flat-fading model (constant within the transmission frame) is adopted for all channels. As discussed in \cite{wu2019towards, guan2020intelligent,zheng2019intelligent}, by applying various channel acquisition methods, we can acquire all channel information, and hence here for the current study, we also assume that the Channel State Information (CSI) of all channels are perfectly known.
Linear transmit precoding is considered at the base station similar to \cite{wu2019intelligent}, and each user served by the base station is assigned with one dedicated beamforming vector. To further enhance the physical layer security, additional AN is also adopted. Thus, the signal transmitted from the base station to the $i$-th user can be described as:


\begin{equation}\label{signal}
\boldsymbol{x}_i=\boldsymbol{\omega}_is_i+\boldsymbol{z}_i, i=1,...U,
\end{equation}
where $\boldsymbol{\omega}_i\in\mathbb{C}^{M\times 1}$ is the beamforming vector for the $i$-th user, $s_i$ is the corresponding transmitted data, and $\boldsymbol{z}_i\in\mathbb{C}^{M\times 1}$ is an AN vector.

Since multiple IRSs have been deployed in the system, each legitimate user can be served by a selected IRS to receive tuned signal, which is effective especially when there exists an obstacle and no Light-of-Sight (LoS) channel between the base station and a user.
Let $\alpha_{i,k}\in\{0,1\}$ denote the IRS selection for the $i$-th user, i.e., the $i$-th user can receive reflecting signal through the $k$-th IRS if $\alpha_{i,k} = 1$.
Meanwhile, let $\boldsymbol{\Theta}_k={\rm diag}(A_{k,1}e^{j\theta_{k,1}}, ..., A_{k,N}e^{j\theta_{k,N}})\in\mathbb{C}^{N\times N}$ denote the diagonal phase-shifting matrix of the $k$-th IRS, while $A_{k,n}\in[0,1]$ and $\theta_{k,n}\in[0,2\pi)$ denote the amplitude reflection coefficient and the phase shift of the $n$-th element on the $k$-th IRS. In practice, each element of an IRS is usually designed to maximize the
signal reflection \cite{wu2019intelligent}. Thus, we set $A_{k,n}=1$ in this paper.
In this case, for the $i$-th user, the received signal from base station and IRSs can be represented by:
\begin{align}\label{receivedSignal}
\boldsymbol{y}_i=&(\sum\limits_{k=1}^K\alpha_{i,k}\boldsymbol{g}_{i,k}^H\boldsymbol{\Theta}_k\boldsymbol{G}^H_k+\boldsymbol{h}_i^H)(\boldsymbol{\omega}_i s_i+\boldsymbol{z}_i)+\notag\\
&\sum\limits_{j\neq i}(\sum\limits_{k=1}^K\alpha_{j,k}\boldsymbol{g}_{i,k}^H\boldsymbol{\Theta}_k\boldsymbol{G}^H_k+\boldsymbol{h}_i^H)(\boldsymbol{\omega}_j s_i+\boldsymbol{z}_j)+n_0,
\end{align}
where $n_0\in \mathcal{CN}(0, \sigma^2)$ is the complex Additive White Gaussian Noise (AWGN).
For an eavesdropper, the received signal can be represented by:
\begin{align}\label{receivedSignalEva}
\boldsymbol{y}^e_i=&(\sum\limits_{k=1}^K\alpha_{i,k}\boldsymbol{g}_{e,k}^H\boldsymbol{\Theta}_k\boldsymbol{G}^H_k+\boldsymbol{h}^H_e)(\boldsymbol{\omega}_i s_i+\boldsymbol{z}_i)\notag\\
&\sum\limits_{j\neq i}(\sum\limits_{k=1}^K\alpha_{j,k}\boldsymbol{g}_{e,k}^H\boldsymbol{\Theta}_k\boldsymbol{G}^H_k+\boldsymbol{h}_e^H)(\boldsymbol{\omega}_j s_i+\boldsymbol{z}_j)+n_0.
\end{align}
For notational simplicity, let $\boldsymbol{\hat{D}}_{ij}=\sum\limits_{k=1}^K\alpha_{j,k}\boldsymbol{g}^H_{i,k}\boldsymbol{\Theta}_k\boldsymbol{G}^H_k+\boldsymbol{h}^H_{i}\in\mathbb{C}^{1\times M}$,
$\boldsymbol{D}_{e,i}=\sum\limits_{k=1}^K\alpha_{i,k}\boldsymbol{g}^H_{e,k}\boldsymbol{\Theta}_k\boldsymbol{G}^H_k+\boldsymbol{h}^H_e\in\mathbb{C}^{1\times M}$.
Accordingly, the Signal-to-Noise Ratio (SINR) of received signal at the $i$-th user can be calculated by:
\begin{equation}\label{SINR_user}
SINR_i=\frac{|(\sum\limits_{k=1}^K\alpha_{i,k}\boldsymbol{g}_{i,k}^H\boldsymbol{\Theta}_k\boldsymbol{G}^H_k+\boldsymbol{h}_i^H)\boldsymbol{\omega}_i |^2}{\sum\limits_{j\neq i}|(\boldsymbol{\hat{D}}_{ij}\boldsymbol{\omega}_j |^2+\sum\limits_{j\in\mathcal{U}}|\boldsymbol{\hat{D}}_{ij}\boldsymbol{z}_j|^2+N_0},
\end{equation}
where $N_0$ is the power of AWGN. Similarly, the SINR of the $i$-th user's signal at the eavesdropper can be calculated by:
\begin{equation}\label{SINR_eva}
SINR^e_i=\frac{|(\sum\limits_{k=1}^K\alpha_{i,k}\boldsymbol{g}_{e,k}^H\boldsymbol{\Theta}_k\boldsymbol{G}^H_k+\boldsymbol{h}_e^H)\boldsymbol{\omega}_i |^2}{\sum\limits_{j\neq i}|\boldsymbol{D}_{e,j}\boldsymbol{\omega}_j |^2+\sum\limits_{j\in\mathcal{U}}|\boldsymbol{D}_{e,j}\boldsymbol{z}_j|^2+N_0}.
\end{equation}

\section{Problem Formulation and Solution}
Considering the security requirement for each legitimate user in the system, we want to guarantee the worst performance of all legitimate users in case an eavesdropper might wiretap/intercept too much useful information from a certain user. Thus, in this paper, we aim to maximize the minimum achievable secrecy rate of legitimate users in the system. By jointly configuring the beamforming matrix $\boldsymbol{\bar{\omega}}=[\boldsymbol{\omega}_1, \boldsymbol{\omega}_2 ..., \boldsymbol{\omega}_I]$ and AN matrix $\boldsymbol{\bar{z}}=[\boldsymbol{z}_1, \boldsymbol{z}_2 ..., \boldsymbol{z}_I]$ at the base station, phase shift matrix $\boldsymbol{\bar{\Theta}}=[\boldsymbol{\Theta}_1, \boldsymbol{\Theta}_2, ..., \boldsymbol{\Theta}_K]$ at IRSs and surface selection matrix $\boldsymbol{\bar{\alpha}}=\begin{bmatrix}
   \boldsymbol{\alpha}_{1,1} & ... & \boldsymbol{\alpha}_{1,K} \\
   ... & ... & ... \\
   \boldsymbol{\alpha}_{I,1} & ... & \boldsymbol{\alpha}_{I,K}
  \end{bmatrix}$ between users and IRSs, the optimization problem can be formulated as:
\begin{equation}
\label{problem1}
\textbf{Problem ~1}:~~~\max\limits_{\boldsymbol{\bar{\omega}},\boldsymbol{\bar{z}},\boldsymbol{\bar{\Theta}},\boldsymbol{\bar{\alpha}}}
~\min\limits_i~~~ [R_i^u-R_i^e]^+
\end{equation}
\begin{align}
s.t.~~~
&||\boldsymbol{\omega}_i||^2+||\boldsymbol{z}_i||^2\leq P_{max},~\forall i\in\mathcal{U}, \tag{C1}\\
&0\leq\theta_{k,n}\leq2\pi,~ k\in [1,K],\forall n\in[1,N], \tag{C2}\\
&\sum\limits_k\alpha_{i,k}= 1, ~\alpha_{i,k}\in\{0,1\}, ~\forall i\in\mathcal{U}, \tag{C3}
\end{align}
where (C1) represents the transmission power constraint, (C2) implies the unit modulus for each element, i.e., $|e^{j\theta_{k,n}}|=1$, and (C3) indicates that each user should be served by one IRS in the system.
Considering the SINR expression in (\ref{SINR_user}), (\ref{SINR_eva}) and applying the Shannon capacity theorem, the achievable secrecy rate (bits/s/Hz) in (\ref{problem1}) can be calculated by:
\begin{align}\label{achievable_secrecy_rate}
&R^u_i-R^e_i=\log_2(1+\frac{|(\sum\limits_{k=1}^K\alpha_{i,k}\boldsymbol{g}^H_{i,k}\boldsymbol{\Theta}_kG^H_k+\boldsymbol{h}^H_{i})\boldsymbol{\omega}_i|^2}{\sum\limits_{j\neq i}|(\boldsymbol{\hat{D}}_{ij}\boldsymbol{\omega}_j |^2+\sum\limits_{j\in\mathcal{U}}|\boldsymbol{\hat{D}}_{ij}\boldsymbol{z}_j|^2+N_0})\notag\\
&-\log_2(1+\frac{|(\sum\limits_{k=1}^K\alpha_{i,k}\boldsymbol{g}^H_{e,k}\boldsymbol{\Theta}_kG^H_k+\boldsymbol{h}^H_{e})\boldsymbol{\omega}_i|^2}{\sum\limits_{j\neq i}|(\boldsymbol{D}_{e,j}\boldsymbol{\omega}_j |^2+\sum\limits_{j\in\mathcal{U}}|\boldsymbol{D}_{e,j}\boldsymbol{z}_j|^2+N_0}).
\end{align}

It is intuitive that variables $\boldsymbol{\bar{\omega}}$, $\boldsymbol{\bar{z}}$, $\boldsymbol{\bar{\Theta}}$ and $\boldsymbol{\bar{\alpha}}$ in \textbf{Problem 1} are coupled, which makes \textbf{Problem 1} hard to solve.
However, if only one variable is considered, the original problem becomes solvable.
Inspired by the alternating optimization approaches in \cite{guan2020intelligent, yu2019enabling, dong2020secure, lyu2020irs, xu2019resource, wu2019intelligent,li2020energy}, we adopt Block Coordinate Descent (BCD) method to decouple variables and obtain the sub-optimal solution efficiently.
To optimize a multi-variable objective in BCD method, we optimize the objective in terms of one of the coordinate blocks while the other blocks are fixed at each iteration.
Thus, \textbf{Problem 1} is divided into three sub-problems and each sub-problem is solved iteratively as descried in \textbf{Algorithm \ref{alg:AlternatingAlgorithm}}. For each sub-problem, we utilize SDR and SCA to convert the original problem into a convex problem. The detailed solving process of each sub-problem is descried in the following sub-sections.

\begin{algorithm}[t]
\caption{BCD-based Algorithm}
\label{alg:AlternatingAlgorithm}
\KwIn{Number of elements $N$, number of antennas $M$, number of surfaces $K$;}
\KwOut{Beamforming vector $\boldsymbol{\bar{\omega}}$, AN vector $\boldsymbol{\bar{z}}$, phase -shift matrix $\boldsymbol{\bar{\Theta}}$ and IRS selection vector $\bar{\alpha}$;}
\textbf{Initialize:}\\
{
\begin{itemize}
    \item Initialize $\boldsymbol{\bar{\omega}}^{(0)}$, $\boldsymbol{\bar{z}}^{(0)}$, $\boldsymbol{\bar{\Theta}}^{(0)}$ and $\boldsymbol{\bar{\alpha}}^{(0)}$;
    \item $t=0$, $\Delta^{(t)}=Intmax$;
\end{itemize}
}
\While{$\Delta^{(t)}<\delta$}{
Solve each sub-problem to find solution for $\boldsymbol{\bar{\omega}}^{(t+1)}$, $\boldsymbol{\bar{z}}^{(t+1)}$, $\boldsymbol{\bar{\Theta}}^{(t+1)}$ and $\boldsymbol{\bar{\alpha}}^{(t+1)}$  for given $\boldsymbol{\bar{\omega}}^{(t)}$, $\boldsymbol{\bar{z}}^{(t)}$, $\boldsymbol{
\bar{\Theta}}^{(t)}$ and $\boldsymbol{\bar{\alpha}}^{(t)}$, respectively;\\
Calculate $\rho^{(t+1)}=\min\limits_i~~~ [R_i^u-R_i^e]$;\\
Update $t=t+1$ and $\Delta^{(t)}=\rho^{(t+1)}-\rho^{(t)}$;\\
}
\end{algorithm}

\begin{strip}
\hrulefill
\begin{align}
&\bigtriangledown_{\boldsymbol{W}_i}F_i^3(\boldsymbol{W}_i,\boldsymbol{Z}_i)=0,~\bigtriangledown_{\boldsymbol{Z}_i}F_i^3(\boldsymbol{W}_i,\boldsymbol{Z}_i)=\frac{1}{\rm ln2}\frac{(\boldsymbol{\hat{D}}^H_{i,j}\boldsymbol{\hat{D}}_{i,j})^H}{\sum\limits_{j\neq i}({\rm Tr}(\boldsymbol{W}_j\boldsymbol{\hat{D}}^H_{i,j}\boldsymbol{\hat{D}}_{i,j})+\sum\limits_{j\in\mathcal{U}}{\rm Tr}(\boldsymbol{Z}_j\boldsymbol{\hat{D}}^H_{i,j}\boldsymbol{\hat{D}}_{i,j})+N_0},\label{gradientF3}\\
&\bigtriangledown_{\boldsymbol{W}_i}F_i^4(\boldsymbol{W}_i,\boldsymbol{Z}_i)=\frac{1}{\rm ln2}\frac{(\boldsymbol{D}^H_{e,j}\boldsymbol{D}_{e,j})^H}{{\rm Tr}(\boldsymbol{W}_i\boldsymbol{D}^H_{e,i}\boldsymbol{D}_{e,i})+\sum\limits_{j\neq i}({\rm Tr}(\boldsymbol{W}_j\boldsymbol{D}^H_{e,j}\boldsymbol{D}_{e,j})+\sum\limits_{j\in\mathcal{U}}{\rm Tr}(\boldsymbol{Z}_j\boldsymbol{D}^H_{e,j}\boldsymbol{D}_{e,j})+N_0},\label{gradientF4W}\\
&\bigtriangledown_{\boldsymbol{Z}_i}F_i^4(\boldsymbol{W}_i,\boldsymbol{Z}_i)=\frac{1}{\rm ln2}\frac{(\boldsymbol{D}^H_{e,j}\boldsymbol{D}_{e,j})^H}{{\rm Tr}(\boldsymbol{W}_i\boldsymbol{D}^H_{e,i}\boldsymbol{D}_{e,i})+\sum\limits_{j\neq i}({\rm Tr}(\boldsymbol{W}_j\boldsymbol{D}^H_{e,j}\boldsymbol{D}_{e,j})+\sum\limits_{j\in\mathcal{U}}{\rm Tr}(\boldsymbol{Z}_j\boldsymbol{D}^H_{e,j}\boldsymbol{D}_{e,j})+N_0}.\label{gradientF4Z}
\end{align}
\hrulefill
\end{strip}
\subsection{Sub-Problem for Beamforming and AN}
At first, beamforming and AN matrices are considered to be solved.
For given phase shift operation $\boldsymbol{\bar{\Theta}}$ and surface matching $\boldsymbol{\bar{\alpha}}$, with $[x]^+=\max\{0,x\}$, we can rewrite \textbf{Problem 1} as:
\begin{equation}
\label{problem2a}
\textbf{Problem ~2a}:~~~\max\limits_{\boldsymbol{\bar{\omega}},\boldsymbol{\bar{z}}}~\min\limits_i~~~ [R_i^u-R_i^e]^+
\end{equation}
\begin{align}
s.t.~~~
&||\boldsymbol{\omega}_i||^2+||\boldsymbol{z}_i||^2\leq P_{max},~\forall i\in\mathcal{U}. \tag{C1}
\end{align}
Due to the max function in (\ref{problem2a}), we can rewrite the objective as
\begin{equation}\label{piecewise}
[R_i^u-R_i^e]^+=\left\{
\begin{aligned}
&0,\boldsymbol{\omega}_i,\boldsymbol{z}_i\in\mathcal{A},\\
&R_i^u-R_i^e,\boldsymbol{\omega}_i,\boldsymbol{z}_i\in\mathcal{A}^+,\\
\end{aligned}
\right.
\end{equation}
where $\mathcal{A}$ and $\mathcal{A}^+$ denote the solution space for non-positive and positive values, respectively. Once $\mathcal{A}^+$ is non-empty, the optimal solution must satisfy $(\boldsymbol{\omega}_i^*,\boldsymbol{z}_i^*)\in\mathcal{A}^+$. In this case, we can rewrite (\ref{problem2a}) as $R_i^u-R_i^e$ when $\mathcal{A}^+\neq\emptyset$.

To solve this sub-problem for beamforming and AN, we plan to apply SDR in the next. So we start to reformulate the objective with some mathematical transformations.
Let $\boldsymbol{W}_i=\boldsymbol{\omega}_i\boldsymbol{\omega}_i^H\in\mathbb{C}^{M\times M}$, $\boldsymbol{Z}_i=\boldsymbol{z}_i\boldsymbol{z}_i^H\in\mathbb{C}^{M\times M}$, $\boldsymbol{D}_{i}=\sum\limits_{k=1}^K\alpha_{i,k}\boldsymbol{g}^H_{i,k}\boldsymbol{\Theta}_k\boldsymbol{G}^H_k+\boldsymbol{h}^H_{i}\in\mathbb{C}^{1\times M}$,
$\boldsymbol{\hat{D}}_{ij}=\sum\limits_{k=1}^K\alpha_{j,k}\boldsymbol{g}^H_{i,k}\boldsymbol{\Theta}_k\boldsymbol{G}^H_k+\boldsymbol{h}^H_{i}\in\mathbb{C}^{1\times M}$,
$\boldsymbol{D}_{e,i}=\sum\limits_{k=1}^K\alpha_{i,k}\boldsymbol{g}^H_{e,k}\boldsymbol{\Theta}_k\boldsymbol{G}^H_k+\boldsymbol{h}^H_e\in\mathbb{C}^{M_e\times M}$.
Then, the achievable secrecy rate can be reformulated as:
{\footnotesize
\begin{align}
&R^u_i-R^e_i=\log_2(1+\frac{{\rm Tr}(\boldsymbol{W}_i\boldsymbol{D}^H_{i}\boldsymbol{D}_{i})}{\sum\limits_{j\neq i}(({\rm Tr}(\boldsymbol{W}_j\boldsymbol{\hat{D}}^H_{i,j}\boldsymbol{\hat{D}}_{i,j})+\sum\limits_{j\in\mathcal{U}}{\rm Tr}(\boldsymbol{Z}_j\boldsymbol{\hat{D}}^H_{i,j}\boldsymbol{\hat{D}}_{i,j})+N_0})\notag\\
&-\log_2(1+\frac{{\rm Tr}(\boldsymbol{W}_i\boldsymbol{D}^H_{e,i}\boldsymbol{D}_{e,i})}{\sum\limits_{j\neq i}({\rm Tr}(\boldsymbol{W}_j\boldsymbol{D}^H_{e,j}\boldsymbol{D}_{e,j})+\sum\limits_{j\in\mathcal{U}}{\rm Tr}(\boldsymbol{Z}_j\boldsymbol{D}^H_{e,j}\boldsymbol{D}_{e,j})+N_0}),\notag\\
&=\log_2(\frac{{\rm Tr}(\boldsymbol{W}_i\boldsymbol{D}^H_{i}\boldsymbol{D}_{i})+\sum\limits_{j\neq i}{\rm Tr}(\boldsymbol{W}_j\boldsymbol{\hat{D}}^H_{ij}\boldsymbol{\hat{D}}_{ij})+\sum\limits_{j\in\mathcal{U}}{\rm Tr}(\boldsymbol{Z}_j\boldsymbol{\hat{D}}^H_{ij}\boldsymbol{\hat{D}}_{ij})+N_0}{\sum\limits_{j\neq i}{\rm Tr}(\boldsymbol{W}_j\boldsymbol{\hat{D}}^H_{ij}\boldsymbol{\hat{D}}_{ij})+\sum\limits_{j\in\mathcal{U}}{\rm Tr}(\boldsymbol{Z}_j\boldsymbol{\hat{D}}^H_{ij}\boldsymbol{\hat{D}}_{ij})+N_0}\notag\\
&\cdot\frac{\sum\limits_{j\neq i}{\rm Tr}(\boldsymbol{W}_j\boldsymbol{D}^H_{e,j}\boldsymbol{D}_{e,j})+\sum\limits_{j\in\mathcal{U}}{\rm Tr}(\boldsymbol{Z}_j\boldsymbol{D}^H_{e,j}\boldsymbol{D}_{e,j})+N_0}{{\rm Tr}(\boldsymbol{W}_i\boldsymbol{D}^H_{e,i}\boldsymbol{D}_{e,i})+\sum\limits_{j\neq i}{\rm Tr}(\boldsymbol{W}_j\boldsymbol{D}^H_{e,j}\boldsymbol{D}_{e,j})+\sum\limits_{j\in\mathcal{U}}{\rm Tr}(\boldsymbol{Z}_j\boldsymbol{D}^H_{e,j}\boldsymbol{D}_{e,j})+N_0}),\notag\\
&=F^1_i+F^2_i-F^3_i-F^4_i,\label{secrecyRate_W}
\end{align}
}
where $F^1_i$, $F^2_i$, $F^3_i$ and $F^4_i$ are represented by:
{\footnotesize
\begin{align}
&F_i^1=\log_2({\rm Tr}(\boldsymbol{W}_i\boldsymbol{D}^H_{i}\boldsymbol{D}_{i})\notag\\
&~~~~+\sum\limits_{j\neq i}({\rm Tr}(\boldsymbol{W}_j\boldsymbol{\hat{D}}^H_{i,j}\boldsymbol{\hat{D}}_{i,j})+\sum\limits_{j\in\mathcal{U}}{\rm Tr}(\boldsymbol{Z}_j\boldsymbol{\hat{D}}^H_{i,j}\boldsymbol{\hat{D}}_{i,j})+N_0),\label{F1}\\
&F_i^2=\log_2(\sum\limits_{j\neq i}{\rm Tr}(\boldsymbol{W}_j\boldsymbol{D}^H_{e,j}\boldsymbol{D}_{e,j})+\sum\limits_{j\in\mathcal{U}}{\rm Tr}(\boldsymbol{Z}_j\boldsymbol{D}^H_{e,j}\boldsymbol{D}_{e,j})+N_0),\label{F2}\\
&F_i^3=\log_2(\sum\limits_{j\neq i}({\rm Tr}(\boldsymbol{W}_j\boldsymbol{\hat{D}}^H_{i,j}\boldsymbol{\hat{D}}_{i,j})+\sum\limits_{j\in\mathcal{U}}{\rm Tr}(\boldsymbol{Z}_j\boldsymbol{\hat{D}}^H_{i,j}\boldsymbol{\hat{D}}_{i,j})+N_0),\label{F3}\\
&F_i^4=\log_2({\rm Tr}(\boldsymbol{W}_i\boldsymbol{D}^H_{e,i}\boldsymbol{D}_{e,i})\notag\\
&~~~~+\sum\limits_{j\neq i}({\rm Tr}(\boldsymbol{W}_j\boldsymbol{D}^H_{e,j}\boldsymbol{D}_{e,j})+\sum\limits_{j\in\mathcal{U}}{\rm Tr}(\boldsymbol{Z}_j\boldsymbol{D}^H_{e,j}\boldsymbol{D}_{e,j})+N_0).\label{F4}
\end{align}
}

However, the secrecy rate $R^u_i-R^e_i$ in (\ref{secrecyRate_W}) is still in the form of Difference of Convex (DC) functions. To solve the DC problem in (\ref{secrecyRate_W}), we adopt SCA method \cite{razaviyayn2014successive,sun2017optimal,xu2019resource,alvarado2014new} to obtain a convex upper bound for the DC objective in an iterative manner.
At first, we construct global upper bound of $F_i^3$ and $F_i^4$, respectively.
For any feasible solution $(\boldsymbol{W}_i^{(t)}, \boldsymbol{Z}_i^{(t)})$, the differentiable convex
functions $F_i^3(\boldsymbol{W}_i,\boldsymbol{Z}_i)$ and $F_i^4(\boldsymbol{W}_i,\boldsymbol{Z}_i)$ satisfy the following inequalities\footnote{Since $F_i^3$ and $F_i^4$ are concave functions, according to the definition of concave function, we have $(1-\lambda) f(x)+\lambda f(y)\leq f((1-\lambda)x+\lambda y)$. In this case, we can construct global upper bound $f(y)\leq \frac{f((1-\lambda)x+\lambda y)- (1-\lambda) f(x)}{\lambda}=f(x)+\frac{f(x+\lambda (y-x))- f(x) }{\lambda}\rightarrow f(x)+\bigtriangledown f(x)(y-x)$ as $\lambda\rightarrow0$ \citep[Proposition 1.8]{simchowitz2018course}.}:
\begin{align}
F_i^3(\boldsymbol{W}_i,\boldsymbol{Z}_i)&\leq F^3_i(\boldsymbol{W}_i^{(t)},\boldsymbol{Z}_i^{(t)})\notag\\
&+{\rm Tr}(\bigtriangledown_{\boldsymbol{W}_i}F^3_i(\boldsymbol{W}_i^{(t)},\boldsymbol{Z}_i^{(t)})^H(\boldsymbol{W}_i-\boldsymbol{W}_i^{(t)}))\notag\\
&+{\rm Tr}(\bigtriangledown_{\boldsymbol{Z}_i}F^3_i(\boldsymbol{W}_i^{(t)},\boldsymbol{Z}_i^{(t)})^H(\boldsymbol{Z}_i-\boldsymbol{Z}_i^{(t)}))\notag\\
&=\widetilde{F}_i^3(\boldsymbol{W}_i,\boldsymbol{Z}_i,\boldsymbol{W}_i^{(t)},\boldsymbol{Z}_i^{(t)}),\label{F3_SCA}\\
F_i^4(\boldsymbol{W}_i,\boldsymbol{Z}_i)&\leq F^4_i(\boldsymbol{W}_i^{(t)},\boldsymbol{Z}_i^{(t)})\notag\\
&+{\rm Tr}(\bigtriangledown_{\boldsymbol{W}_i}F^4_i(\boldsymbol{W}_i^{(t)},\boldsymbol{Z}_i^{(t)})^H(\boldsymbol{W}_i-\boldsymbol{W}_i^{(t)}))\notag\\
&+{\rm Tr}(\bigtriangledown_{\boldsymbol{Z}_i}F^4_i(\boldsymbol{W}_i^{(t)},\boldsymbol{Z}_i^{(t)})^H(\boldsymbol{Z}_i-\boldsymbol{Z}_i^{(t)}))\notag\\
&=\widetilde{F}_i^4(\boldsymbol{W}_i,\boldsymbol{Z}_i,\boldsymbol{W}_i^{(t)},\boldsymbol{Z}_i^{(t)}),\label{F4_SCA}
\end{align}
where the right hand side terms in (\ref{F3_SCA}) and (\ref{F4_SCA}) are global upper bound of $F_i^1$ and $F_i^2$ by using first-order Taylor approximation, respectively.
The gradients of functions $F_i^3$ and $F_i^4$ with respect to $\boldsymbol{W}_i$ and $\boldsymbol{Z}_i$ are given in (\ref{gradientF3})-(\ref{gradientF4Z}).
Hence, a convex lower bound of objective function in (\ref{secrecyRate_W}) can be obtained as  $R^u_i-R^e_i=F^1_i+F^2_i-\widetilde{F}^3_i-\widetilde{F}^4_i$.
Let $f_i(\boldsymbol{W}_i, \boldsymbol{Z}_i)=F^1_i+F^2_i-F^3_i-F^4_i$ and $g_i(\boldsymbol{W}_i, \boldsymbol{Z}_i)=F^1_i+F^2_i-\widetilde{F}^3_i-\widetilde{F}^4_i$.
Since $f_i(\boldsymbol{W}_i, \boldsymbol{Z}_i)\geq g_i(\boldsymbol{W}_i, \boldsymbol{Z}_i)$ according to (\ref{F3_SCA}) and (\ref{F4_SCA}), as long as we guarantee $g_i(\boldsymbol{W}_i, \boldsymbol{Z}_i)\geq0$, $f_i(\boldsymbol{W}_i, \boldsymbol{Z}_i)>0$ must be satisfied.

After deploying SCA, the objective function becomes convex. In order to further solve the max-min problem, we also introduce an auxiliary variable $x$ into the formulation.
By doing so, the original \textbf{Problem 2a} can be transformed to:
\begin{equation}
\label{problem2b}
\textbf{Problem ~2b}:~~~\max\limits_{x, \bar{\boldsymbol{W}},\bar{\boldsymbol{Z}}}~~~ x
\end{equation}

\begin{align}
s.t.~~~
&{\rm Tr}(\boldsymbol{W}_i)+{\rm Tr}(\boldsymbol{Z}_i)\leq P_{max},~\forall i\in\mathcal{U}, \tag{C1}\label{C1}\\
&0\leq x\leq F^1_i+F^2_i-\widetilde{F}^3_i-\widetilde{F}^4_i,\forall i\in\mathcal{U},\tag{C4}\label{C4}\\
&{\rm Rank}(\boldsymbol{W}_i)=1,~{\rm Rank}(\boldsymbol{Z}_i)=1,~\forall i\in\mathcal{U},\tag{C5}\\
&\boldsymbol{W}_i\succeq \boldsymbol{0},~\boldsymbol{Z}_i\succeq \boldsymbol{0},~\forall i\in\mathcal{U},\tag{C6}\label{C6}
\end{align}
where $\bar{\boldsymbol{W}}=[\boldsymbol{W}_1,\boldsymbol{W}_2,...,\boldsymbol{W}_I],\bar{\boldsymbol{Z}}=[\boldsymbol{Z}_1,\boldsymbol{Z}_2,...,\boldsymbol{Z}_I]\in\mathbb{C}^{IM\times IM}$.
Since constraint (C5) is non-convex, we drop this rank-1 constraint by applying SDR.
If the obtained solution $(\boldsymbol{W}_i^{(t)}, \boldsymbol{Z}_i^{(t)})$ are of rank-1, they can be written as $\boldsymbol{W}_i^{(t)}=\boldsymbol{\omega}_i\boldsymbol{\omega}_i^H$ and $\boldsymbol{Z}_i^{(t)}=\boldsymbol{z}_i\boldsymbol{z}_i^H$, then the optimal beamforming vector $\boldsymbol{\omega}_i$ and AN $\boldsymbol{z}_i$ can be obtained by applying
eigenvalue decomposition. Otherwise, we can adopt Gaussian Randomization to recover $\boldsymbol{\omega}_i$ and $\boldsymbol{z}_i$ approximately from higher rank solution $(\boldsymbol{W}_i^{(t)}, \boldsymbol{Z}_i^{(t)})$ \cite{luo2010semidefinite,ma2010semidefinite,ma2012semidefinite}.
In this case, \textbf{Problem 2b} becomes a convex optimization problem.
In Algorithm \ref{alg:SCA}, \textbf{Problem 2b} can be efficiently solved at each iteration by using convex optimization solvers, e.g., SeduMi and CVX \cite{sedumi, grant2014cvx}.
In the following, we prove that SCA-based approach in Algorithm \ref{alg:SCA} can reach the optimal solution at each iteration.

\newtheorem*{Proposition 1}{Proposition 1}
\begin{Proposition 1}
Algorithm \ref{alg:SCA} generates a sequence of non-decreasing feasible solutions that converge to a point $(\bar{\boldsymbol{W}}^{*},\bar{\boldsymbol{Z}}^{*})$ satisfying the KKT conditions of the original problems in (\ref{problem2a}).
\end{Proposition 1}
\begin{proof}
For notational convenience
let $f_i(\boldsymbol{W}_i, \boldsymbol{Z}_i)=F^1_i+F^2_i-F^3_i-F^4_i$ and $g_i(\boldsymbol{W}_i, \boldsymbol{Z}_i)=F^1_i+F^2_i-\widetilde{F}^3_i-\widetilde{F}^4_i$.
The constraint (\ref{C4}) can be rewritten as $x\leq\min\limits_{i}\{g_i(\boldsymbol{W}_i, \boldsymbol{Z}_i)\}$.

According to (\ref{F3_SCA}) and (\ref{F4_SCA}), we can obtain $\max\limits_{i}\{f_i(\boldsymbol{W}^{(t)}_i, \boldsymbol{Z}^{(t)}_i)\}\geq \max\limits_{i}\{g_i(\boldsymbol{W}^{(t)}_i, \boldsymbol{Z}^{(t)}_i)\},\forall \bar{\boldsymbol{W}},\bar{\boldsymbol{Z}}\in\mathbb{C}^{IM\times IM}$.
Since constraints (\ref{C1}), (\ref{C4}) and (\ref{C6}) are always satisfied, the optimal solution $(\bar{\boldsymbol{W}}^{(t)}, \bar{\boldsymbol{Z}}^{(t)})$ of the approximated problem (\ref{problem2b}) at the $t$-th iteration always belongs to the feasible set of the original problem (\ref{problem2a})
At each iteration, it follows that \cite{nasir2015joint,wang2012successive}:
\begin{align}
\max\limits_{i}\{f_i(\boldsymbol{W}^{(t)}_i, \boldsymbol{Z}^{(t)}_i)\}&\geq \max\limits_{i}\{g_i(\boldsymbol{W}^{(t)}_i, \boldsymbol{Z}^{(t)}_i)\}\notag\\
&=\min\limits_{\bar{\boldsymbol{W}},\bar{\boldsymbol{Z}}}~\max\limits_{i}\{g_i(\boldsymbol{W}_i, \boldsymbol{Z}_i)\}\notag\\
&\geq\max\limits_{i}\{g_i(\boldsymbol{W}^{(t-1)}_i, \boldsymbol{Z}^{(t-1)}_i)\}\notag\\
&= \max\limits_{i}\{f_i(\boldsymbol{W}^{(t-1)}_i, \boldsymbol{Z}^{(t-1)}_i)\},\notag
\end{align}
where the second inequality holds because $(\bar{\boldsymbol{W}}^{(t)}, \bar{\boldsymbol{Z}}^{(t)})$ is the global optimum of (\ref{problem2b}) at the $t$-th iteration, and the last equality holds because $g_i(\boldsymbol{W}^{(t-1)}_i, \boldsymbol{Z}^{(t-1)}_i)=f_i(\boldsymbol{W}^{(t-1)}_i, \boldsymbol{Z}^{(t-1)}_i)$.
This means that $\{\max\limits_{i}\{f_i(\boldsymbol{W}^{(t)}_i, \boldsymbol{Z}^{(t)}_i)\}|t=0,1,...\}$ is a monotonically increasing sequence. As the actual objective value in (\ref{problem2b}) is nondecreasing after every iteration, Algorithm \ref{alg:SCA} will eventually converge to a point $(\bar{\boldsymbol{W}}^{*},\bar{\boldsymbol{Z}}^{*})$ as $t$ increases.

Next, we prove that $(\bar{\boldsymbol{W}}^{*},\bar{\boldsymbol{Z}}^{*})$ satisfies the KKT conditions of the original problem. From (\ref{problem2b}), the optimal solution can be found when $x=\min\limits_{i}\{g_i(\boldsymbol{W}_i, \boldsymbol{Z}_i)\}$, thus, \textbf{Problem 2b} can be rewritten as:
\begin{equation}
\label{problem2b-change}
\max\limits_{x, \bar{\boldsymbol{W}},\bar{\boldsymbol{Z}}}~~~ \min\limits_{i}\{g_i(\boldsymbol{W}_i, \boldsymbol{Z}_i)\}
\end{equation}
\begin{align}
s.t.~~~
&(C1),(C6).\notag
\end{align}
Then, the Lagrangian for (\ref{problem2b-change}) is:
\begin{align}
L&(\bar{\boldsymbol{W}},\bar{\boldsymbol{Z}},\boldsymbol{\mu})=\min\limits_{i}\{g_i(\boldsymbol{W}_i, \boldsymbol{Z}_i)\}+\sum\limits_{i\in\mathcal{U}}\mu_i({\rm Tr}(\boldsymbol{W}_i)+{\rm Tr}(\boldsymbol{Z}_i)),\notag
\end{align}
where $\mu_i$ is the Lagrangian multiplier for each constraint.
Similar to (\ref{problem2b}), by adopting mathematical transformations and introducing auxiliary variable $x$, the Lagrangian for the original problem (\ref{problem2a}) can be written as:
\begin{align}
L'&(\bar{\boldsymbol{W}},\bar{\boldsymbol{Z}},\boldsymbol{\mu})=\min\limits_{i}\{f_i(\boldsymbol{W}_i, \boldsymbol{Z}_i)\}+\sum\limits_{i\in\mathcal{U}}\mu_i({\rm Tr}(\boldsymbol{W}_i)+{\rm Tr}(\boldsymbol{Z}_i)),\notag
\end{align}
For a feasible point $(\bar{\boldsymbol{W}}^{(t-1)},\bar{\boldsymbol{Z}}^{(t-1)})$ obtained from Algorithm \ref{alg:SCA} at the $(t-1)$-th iteration, it is the global optimum for (\ref{problem2b-change}), the KKT conditions of (\ref{problem2b-change}) must be satisfied, i.e., $(\bar{\boldsymbol{W}}^{(t-1)},\bar{\boldsymbol{Z}}^{(t-1)})$ is feasible for (\ref{problem2b-change}) and there exist nonnegative real values $u_i, i\in \mathcal{U}$ satisfying:
\begin{align}
&\bigtriangledown L(\bar{\boldsymbol{W}}^{(t-1)},\bar{\boldsymbol{Z}}^{(t-1)},\boldsymbol{\mu})|_{\boldsymbol{W},\boldsymbol{Z}}=0,\notag\\
&\mu_i({\rm Tr}(\boldsymbol{W}_i^{(t-1)})+{\rm Tr}(\boldsymbol{Z}_i^{(t-1)}))=0,~\forall i\in\mathcal{U}.\notag
\end{align}
Since the gradient of the first-order Taylor approximations $\widetilde{F}_i^3(\boldsymbol{W}_i,\boldsymbol{Z}_i)$ and $\widetilde{F}_i^4(\boldsymbol{W}_i,\boldsymbol{Z}_i)$ are the same as $F_i^3(\boldsymbol{W}_i,\boldsymbol{Z}_i)$ and $F_i^4(\boldsymbol{W}_i,\boldsymbol{Z}_i)$, we can also verify that:
\begin{align}
&\bigtriangledown L'(\bar{\boldsymbol{W}},\bar{\boldsymbol{Z}},\boldsymbol{\mu})|_{\bar{\boldsymbol{W}}=\bar{\boldsymbol{W}}^{(t-1)}}=\bigtriangledown L(\bar{\boldsymbol{W}},\bar{\boldsymbol{Z}},\boldsymbol{\mu})|_{\bar{\boldsymbol{W}}=\bar{\boldsymbol{W}}^{(t-1)}},\notag\\
&\bigtriangledown L'(\bar{\boldsymbol{W}},\bar{\boldsymbol{Z}},\boldsymbol{\mu})|_{\bar{\boldsymbol{Z}}=\bar{\boldsymbol{Z}}^{(t-1)}}=\bigtriangledown L(\bar{\boldsymbol{W}},\bar{\boldsymbol{Z}},\boldsymbol{\mu})|_{\bar{\boldsymbol{Z}}=\bar{\boldsymbol{Z}}^{(t-1)}}.\notag
\end{align}
which implies that $(\bar{\boldsymbol{W}}^{(t-1)},\bar{\boldsymbol{Z}}^{(t-1)})$ satisfies the KKT conditions for (\ref{problem2a}). The results imply that the KKT conditions of the original problem will be satisfied after the series of approximations converges to the point $(\bar{\boldsymbol{W}}^{*},\bar{\boldsymbol{Z}}^{*})$. This completes the proof.

\end{proof}

\begin{algorithm}[t]
\caption{SCA-based Algorithm}
\label{alg:SCA}
\KwIn{Number of elements $N$, number of antennas $M$, number of surfaces $K$;}
\KwOut{Beamforming $\bar{\boldsymbol{W}}^*$, AN $\bar{\boldsymbol{Z}}^*$;}
\textbf{Initialize:}\\
{
\begin{itemize}
    \item Initialize $\bar{\boldsymbol{W}}^{(0)}$, $\bar{\boldsymbol{Z}}^{(0)}$, $t=1$, $\Delta^{(t)}=Intmax$;
\end{itemize}
}
\While{$\Delta^{(t)}<\delta$}{
Solve problem (\ref{problem2b}) to find solution $\bar{\boldsymbol{W}}^{(t)}$, $\bar{\boldsymbol{Z}}^{(t)}$;\\
Update $t=t+1$ and $\Delta^{(t)}=x^{(t+1)}-x^{(t)}$;\\
}
\end{algorithm}


\subsection{Subproblem for Phase Shift}
For given beamforming matrix $\boldsymbol{\bar{\omega}}$, AN matrix $\boldsymbol{\bar{z}}$ and surface selection matrix $\boldsymbol{\bar{\alpha}}$, we can rewrite \textbf{Problem 1} as:
\begin{equation}
\label{problem3a}
\textbf{Problem ~3a}:~~~\max\limits_{\boldsymbol{\bar{\Theta}}}~\min\limits_i~~~ [R_i^u-R_i^e]^+
\end{equation}
\begin{align}
s.t.~~~
&0\leq\theta_{k,n}\leq2\pi,~ k\in [1,K],\forall n\in[1,N]. \tag{C2}
\end{align}

Next, similar to the procedures in the previous section III-A, we also transform the objective function to a solvable convex function by applying SDR and SCA.
Let $\boldsymbol{\mathcal{G}}_{i,k}=\alpha_{i,k}{\rm diag}(\boldsymbol{g}^H_{i,k})\boldsymbol{G}^H_k\in\mathbb{C}^{N\times M}$
\footnote{This is due to $\boldsymbol{A}\cdot {\rm diag}(e^{j\theta_1},e^{j\theta_2},...,e^{j\theta_N}) = [e^{j\theta_1},e^{j\theta_2},...,e^{j\theta_N}]\cdot {\rm diag}(\boldsymbol{A})$ when matrix $\boldsymbol{A}\in\mathbb{C}^{1\times N}$ and ${\rm diag}(e^{j\theta_1},e^{j\theta_2},...,e^{j\theta_N})\in\mathbb{C}^{N\times N}$. Thus, transmitter-IRS-user channel gives $\boldsymbol{g}^H_{i,k}\boldsymbol{\Theta}_kG^H_k=\mu_{k}{\rm diag}(\boldsymbol{g}^H_{i,k})\boldsymbol{G}^H_k$.}, $\boldsymbol{\mathcal{\hat{G}}}_{i,j,k}=\alpha_{j,k}{\rm diag}(\boldsymbol{g}^H_{i,k})\boldsymbol{G}^H_k\in\mathbb{C}^{N\times M}$.
Let $\boldsymbol{\kappa}_{i,k}=\boldsymbol{\mathcal{G}}_{i,k}\boldsymbol{\omega}_i\in \mathbb{C}^{N\times1}$, $\boldsymbol{\hat{\kappa}}_{i,j,k}=\boldsymbol{\mathcal{\hat{G}}}_{i,j,k}\boldsymbol{\omega}_j\in \mathbb{C}^{N\times1}$, $\boldsymbol{\mu}_k=[e^{j\theta_1},e^{j\theta_2},...,e^{j\theta_N}]\in\mathbb{C}^{1\times N}$ and $\mu_{k,n} = e^{j\theta_{k,n}}$. Then, the power of received signal at the $i$-th user in (\ref{SINR_user}) becomes: $$|(\sum\limits_{k=1}^K\alpha_{i,k}\boldsymbol{g}^H_{i,k}\boldsymbol{\Theta}_k\boldsymbol{G}^H_k\boldsymbol{\omega}_i+\boldsymbol{h}^H_{i})\boldsymbol{\omega}_i|^2
=|\sum\limits_{k=1}^K\boldsymbol{\mu}_k\boldsymbol{\kappa}_{i,k}+\boldsymbol{h}^H_i\boldsymbol{\omega}_i|^2.$$
Accordingly, the power of the received signal of the $i$-th user at the eavesdropper in (\ref{SINR_eva}) becomes:
$$|(\sum\limits_{k=1}^K\alpha_{i,k}\boldsymbol{g}^H_{e,k}\boldsymbol{\Theta}_k\boldsymbol{G}^H_k\boldsymbol{\omega}_i+\boldsymbol{h}^H_{i})\boldsymbol{\omega}_i|^2
=|\sum\limits_{k=1}^K\boldsymbol{\mu}_k\boldsymbol{\kappa}_{e,k}+\boldsymbol{h}^H_e\boldsymbol{\omega}_i|^2.$$

Furthermore, let $\boldsymbol{v} = [\boldsymbol{\mu}_1, \boldsymbol{\mu}_2, ..., \boldsymbol{\mu}_K]\in\mathbb{C}^{1\times NK}$, and $\boldsymbol{a}_{i} = [\boldsymbol{\kappa}_{i,1}; \boldsymbol{\kappa}_{i,2}; ...; \boldsymbol{\kappa}_{i,K}]\in\mathbb{C}^{NK\times1}$, $\boldsymbol{\hat{a}}_{i,j} = [\boldsymbol{\hat{\kappa}}_{i,j,1}; \boldsymbol{\hat{\kappa}}_{i,j,2}; ...; \boldsymbol{\hat{\kappa}}_{i,j,K}]\in\mathbb{C}^{NK\times1}$.
Thus, we have $\sum\limits_{k=1}^K\boldsymbol{\mu}_k^H\boldsymbol{\kappa}_{i,k} = \boldsymbol{v}\boldsymbol{a}_{i}$.
Let $b_i = \boldsymbol{h}^H_{i}\boldsymbol{\omega}_{i}$, $\hat{b}_{i,j} = \boldsymbol{h}^H_{i}\boldsymbol{\omega}_{j}$, $\boldsymbol{\mathcal{G}}^e_{i,k}=\alpha_{i,k}{\rm diag}(\boldsymbol{g}^H_{e,k})\boldsymbol{G}_k$, $b^e_{i} = \boldsymbol{h}^H_{e}\boldsymbol{\omega}_{i}$.
Also, let $\boldsymbol{\hat{\kappa}}^{noise}_{i,j,k}=\boldsymbol{\mathcal{\hat{G}}}_{i,j,k}\boldsymbol{z}_i\in \mathbb{C}^{N\times1}$, $\boldsymbol{a}^{noise}_{i,j} = [\boldsymbol{\kappa}^{noise}_{i,j,1}; \boldsymbol{\kappa}^{noise}_{i,j,2}; ...; \boldsymbol{\kappa}^{noise}_{i,j,K}]\in\mathbb{C}^{NK\times1}$, $c_{i,j} = \boldsymbol{h}^H_{i}\boldsymbol{z}_{j}$.
Then, the achievable secrecy rate in (\ref{achievable_secrecy_rate}) can be reformulated as:
{\footnotesize
\begin{align}
\label{secrecyrate_phaseshift}
&R^u_i-R^e_i=\log_2(1+\frac{|\boldsymbol{v}\boldsymbol{a}_{i}+b_{i}|^2}{\sum\limits_{j\neq i}|\boldsymbol{v}\boldsymbol{\hat{a}}_{i,j}+\hat{b}_{i,j}|^2+\sum\limits_{j\in\mathcal{U}}|\boldsymbol{v}\boldsymbol{a}^{noise}_{i,j}+c_{i,j}|^2+N_0})\notag\\
&-\log_2(1+\frac{|\boldsymbol{v}\boldsymbol{a}^e_{i}+\boldsymbol{b}^e_{i}|^2}{\sum\limits_{j\neq i}|\boldsymbol{v}\boldsymbol{a}_{e,j}+\boldsymbol{b}^e_{j}|^2+\sum\limits_{j\in\mathcal{U}}|\boldsymbol{v}\boldsymbol{a}^{noise}_{e,j}+c_{e,j}|^2+N_0}).
\end{align}
}
Note that $|\boldsymbol{v}\boldsymbol{a}_{i}+b_{i}|^2 = \boldsymbol{\widetilde{v}}^H\boldsymbol{R}_{i}\boldsymbol{\widetilde{v}}$, and $ \boldsymbol{\widetilde{v}}^H\boldsymbol{R}_{i}\boldsymbol{\widetilde{v}} = trace(\boldsymbol{R}_{i}\boldsymbol{\widetilde{v}}\boldsymbol{\widetilde{v}}^H)$. Define $\boldsymbol{V} = \boldsymbol{\widetilde{v}}\boldsymbol{\widetilde{v}}^H$, which needs to satisfy $\boldsymbol{V}  \succeq \boldsymbol{0}$ and ${\rm Rank}(\boldsymbol{V})=1$. Note that $\boldsymbol{R}_{i}=[\boldsymbol{a}_{i}\boldsymbol{a}^H_{i}, \boldsymbol{a}_{i}b_i^H;b_i\boldsymbol{a}_{i}^H, 0]\in\mathbb{C}^{NK+1\times NK+1}$, $\boldsymbol{\hat{R}}_{i,j}=[\boldsymbol{\hat{a}}_{i,j}\boldsymbol{\hat{a}}^H_{i,j}, \boldsymbol{\hat{a}}_{i,j}\hat{b}_{i,j}^H;\hat{b}_{i,j}\boldsymbol{\hat{a}}_{i,j}^H, 0]\in\mathbb{C}^{NK+1\times NK+1}$, $\boldsymbol{R}^{noise}_{i,j}=[\boldsymbol{a}^{noise}_{i,j}\boldsymbol{a}^{noise H}_{i,j}, \boldsymbol{a}^{noise}_{i,j}c_{i,j}^H;c_{i,j}\boldsymbol{a}^{noise H}_{i,j}, 0]\in\mathbb{C}^{NK+1\times NK+1}$, $\boldsymbol{\widetilde{v}}=[\boldsymbol{v},\boldsymbol{1}]^H\in\mathbb{C}^{NK+1\times 1}$.
Then (\ref{secrecyrate_phaseshift}) can be further reformulated as:
\begin{align}
\label{rate_phaseshift}
&R^u_i-R^e_i=F_i^1+F_i^2-F_i^3-F_i^4,
\end{align}
where $F^1_i$, $F^2_i$, $F^3_i$ and $F^4_i$ are:
\begin{align}
&F_i^1=\log_2(Tr(\boldsymbol{R}_{i}\boldsymbol{V}) + |b_i|^2\notag
+\sum\limits_{j\neq i}(Tr(\boldsymbol{\hat{R}}_{i,j}\boldsymbol{V})+|\hat{b}_{i,j}|^2)\\
&~~~~~~~~~~~+\sum\limits_{j\in\mathcal{U}}(Tr(\boldsymbol{R}^{noise}_{i,j}\boldsymbol{V})+|c_{i,j}|^2)+N_0),\\
&F_i^2=\log_2(\sum\limits_{j\neq i}(Tr(\boldsymbol{\hat{R}}_{e,j}\boldsymbol{V})+ |\hat{b}_{e,j}|^2)\notag\\
&~~~~~~~~~~~+\sum\limits_{j\in\mathcal{U}}(Tr(\boldsymbol{R}^{noise}_{e,j}\boldsymbol{V})+|c_{e,j}|^2)+N_0),\\
&F_i^3=\log_2(\sum\limits_{j\neq i}(Tr(\boldsymbol{\hat{R}}_{i,j}\boldsymbol{V})+|\hat{b}_{i,j}|^2)\notag\\
&~~~~~~~~~~~+\sum\limits_{j\in\mathcal{U}}(Tr(\boldsymbol{R}^{noise}_{i,j}\boldsymbol{V})+|c_{i,j}|^2)+N_0),\\
&F_i^4=\log_2(Tr(\boldsymbol{R}_{e}\boldsymbol{V}) + |\boldsymbol{b}_i^e|^2\notag
+\sum\limits_{j\neq i}(Tr(\boldsymbol{\hat{R}}_{e,j}\boldsymbol{V})+|\hat{b}_{e,j}|^2)\\
&~~~~~~~~~~~+\sum\limits_{j\in\mathcal{U}}(Tr(\boldsymbol{R}^{noise}_{e,j}\boldsymbol{V})+|c_{e,j}|^2)+N_0).
\end{align}
Similarly, we apply the SDR method to remove rank-one constraint ${\rm Rank}(\boldsymbol{V})=1$ and SCA method to construct global upper bounds of $F^3_i$ and $F^4_i$ and make (\ref{rate_phaseshift}) become convex function:
\begin{align}
F_i^3(\boldsymbol{V})&\leq F^3_i(\boldsymbol{V}^{(t)})+{\rm Tr}(\bigtriangledown_{\boldsymbol{V}}F^3_i(\boldsymbol{V}^{(t)})^H(\boldsymbol{V}-\boldsymbol{V}^{(t)}))\notag\\
&=\widetilde{F}_i^3(\boldsymbol{V},\boldsymbol{V}^{(t)}),\\
F_i^4(\boldsymbol{V})&\leq F^4_i(\boldsymbol{V}^{(t)})+{\rm Tr}(\bigtriangledown_{\boldsymbol{V}}F^4_i(\boldsymbol{V}^{(t)})^H(\boldsymbol{V}-\boldsymbol{V}^{(t)}))\notag\\
&=\widetilde{F}_i^4(\boldsymbol{V},\boldsymbol{V}^{(t)}).
\end{align}
Thus, \textbf{Problem ~3a} is transformed into a convex problem by introducing auxiliary variable $x$:

\begin{equation}
\textbf{Problem ~3b}:~~~\max\limits_{x,\boldsymbol{V}}~~~ x
\end{equation}
\begin{align}
s.t.~~~
&0\leq\theta_{k,n}\leq2\pi,~ k\in [1,K],\forall n\in[1,N], \label{C2}\tag{C2}\\
&0\leq x\leq F^1_i+F^2_i-\widetilde{F}^3_i-\widetilde{F}^4_i,\forall i\in\mathcal{U},\tag{C7}\\
&\boldsymbol{V}\succeq \boldsymbol{0}.\tag{C8}
\end{align}

To restore the desired solution $\boldsymbol{\Theta}={\rm diag}(\boldsymbol{v})$ from the convex Semi-Definite Programming (SDP) solution $\boldsymbol{V}$, eigenvalue decomposition
with Gaussian randomization can be used to obtain a feasible solution based on the higher-rank solution obtained by solving \textbf{Problem ~3b}. Since unit modulus constraint (\ref{C2}) for each element on IRS should be satisfied, the reflection coefficients can be obtained by \cite{wu2019intelligent,guan2020intelligent}:
\begin{equation}
\mu_{k,n}=e^{j\angle(\frac{\mu_{k,n}}{\mu_{NK+1}})}, n = 1,2,...,NK,
\end{equation}
where $\angle(x)$ denotes the phase of $x$ and the obtained solution can satisfy $|\mu_{k,n}|=1$.
\subsection{Subproblem for Surface Selection}
For given beamforming vector $\boldsymbol{\bar{\omega}}$, AN vector $\boldsymbol{\bar{z}}$ and phase shift of IRS $\boldsymbol{\Theta}$, the original problem becomes a 0-1 integer programming problem, and we can rewrite \textbf{Problem ~1} as:
\begin{equation}
\textbf{Problem ~4a}:~~~\max\limits_{\boldsymbol{\alpha}}~ \min\limits_i~~~ [R_i^u-R_i^e]^+
\end{equation}
\begin{align}
s.t.~~~
&\sum\limits_k\alpha_{i,k}= 1, ~\alpha_{i,k}\in\{0,1\}, ~\forall i\in\mathcal{U}. \label{C3}\tag{C3}
\end{align}
At first, according to the constraint described in (\ref{C3}), each user is served by one specific IRS, and thus, we have $\alpha_{i,k}\alpha_{i,k'}=0$ when $k\neq k'$ and $\sum\limits_{k=1}^K\sum\limits_{k'\neq K}\alpha_{i,k}\alpha_{i,k'} = \sum\limits_{k=1}^K\alpha_{i,k}$.
Then, we can simplify the expression in (\ref{SINR_user}) and the power of the received signal at the $i$-th user becomes:
{\small
\begin{align}
&|(\sum\limits_{k=1}^K\alpha_{i,k}\boldsymbol{g}_{i,k}^H\boldsymbol{\Theta}_k\boldsymbol{G}^H_k+\boldsymbol{h}_i^H)\boldsymbol{\omega}_i|^2\notag\\
&=\underbrace{\sum\limits_{k=1}^K\sum\limits_{k'\neq K}\alpha_{i,k}\alpha_{i,k'}(T_{i,k}\boldsymbol{\omega}_i)^HT_{i,k}\boldsymbol{\omega}_i}_{k^2}+(\boldsymbol{h}_i^H\boldsymbol{\omega}_i)^H\boldsymbol{h}_i^H\boldsymbol{\omega}_i\notag\\
&~~~+\sum\limits_{k=1}^K\alpha_{i,k}(T_{i,k}\boldsymbol{\omega}_i)^H\boldsymbol{h}_i^H\boldsymbol{\omega}_i+\sum\limits_{k=1}^K\alpha_{i,k}T_{i,k}\boldsymbol{\omega}_i(\boldsymbol{h}_i^H\boldsymbol{\omega}_i)^H\notag\\
&=\underbrace{\sum\limits_{k=1}^K\alpha_{i,k}(T_{i,k}\boldsymbol{\omega}_i)^HT_{i,k}\boldsymbol{\omega}_i}_k+(\boldsymbol{h}_i^H\boldsymbol{\omega}_i)^H\boldsymbol{h}_i^H\boldsymbol{\omega}_i\notag\\
&~~~+\sum\limits_{k=1}^K\alpha_{i,k}(T_{i,k}\boldsymbol{\omega}_i)^H\boldsymbol{h}_i^H\boldsymbol{\omega}_i+\sum\limits_{k=1}^K\alpha_{i,k}T_{i,k}\boldsymbol{\omega}_i(\boldsymbol{h}_i^H\boldsymbol{\omega}_i)^H\notag\\
&=\sum\limits_{k=1}^K\alpha_{i,k}(\hat{T}^1_{i,i,k}+\hat{T}^2_{i,i,k})+|b_{i}|^2,
\end{align}
}
where $T_{i,k}=\boldsymbol{g}_{i,k}^H\boldsymbol{\Theta}_k\boldsymbol{G}^H_k$, $T_{e,k}=\boldsymbol{g}_{e,k}^H\boldsymbol{\Theta}_k\boldsymbol{G}^H_k$, $\hat{T}^1_{i,j,k}=(T_{i,k}\boldsymbol{\omega}_j)^HT_{i,k}\boldsymbol{\omega}_j$, $\hat{T}^2_{i,j,k}=(T_{i,k}\boldsymbol{\omega}_j)^H\boldsymbol{h}_i^H\boldsymbol{\omega}_j+T_{i,k}\boldsymbol{\omega}_j)(\boldsymbol{h}_i^H\boldsymbol{\omega}_j)^H$.
Similarly, the power of the received signal for the $i$-th user at eavesdropper in (\ref{SINR_eva}) can be expressed as:
{\small
\begin{align}
|(\sum\limits_{k=1}^K\alpha_{i,k}\boldsymbol{g}_{e,k}^H\boldsymbol{\Theta}_k\boldsymbol{G}^H_k+\boldsymbol{h}_e^H)\boldsymbol{\omega}_i|^2
=\sum\limits_{k=1}^K\alpha_{i,k}(\hat{T}^1_{e,i,k}+\hat{T}^2_{e,i,k})+|b_i|^2.
\end{align}
}

Furthermore, let $\hat{TN}^{1}_{i,j,k}=(T_{i,k}\boldsymbol{z}_j)^HT_{i,k}\boldsymbol{z}_j$, $\hat{TN}^{2}_{i,j,k}=(T_{i,k}\boldsymbol{z}_j)^H\boldsymbol{h}_i^H\boldsymbol{z}_j+T_{i,k}\boldsymbol{z}_j)(\boldsymbol{h}_i^H\boldsymbol{z}_j)^H$.
In this case, the achievable secrecy rate in (\ref{achievable_secrecy_rate}) can be reformulated as:
\begin{align}
&R^u_i-R^e_i=F^1_i+F^2_i-F^3_i-F^4_i,\notag
\end{align}
where $F^1_i$, $F^2_i$, $F^3_i$ and $F^4_i$ are represented by:
{\footnotesize
\begin{align}
&F_i^1=\log_2(\sum\limits_{k=1}^K\alpha_{i,k}(\hat{T}^1_{i,i,k}+\hat{T}^2_{i,i,k})+|b_i|^2+\sum\limits_{j\neq i}\sum\limits_{k=1}^K\alpha_{i,k}(\hat{T}^1_{i,j,k}+\hat{T}^2_{i,j,k})\notag\\
&~~~+|\hat{b}_{i,j}|^2+\sum\limits_{j\in\mathcal{U}}\sum\limits_{k=1}^K\alpha_{i,k}(\hat{TN}^{1}_{i,j,k}+\hat{TN}^2_{i,j,k})+|c_{i,j}|^2+N_0),\\
&F_i^2=\log_2(\sum\limits_{j\neq i}\sum\limits_{k=1}^K\alpha_{i,k}(\hat{T}^1_{e,j,k}+\hat{T}^2_{e,j,k})+|\hat{b}_{e,j}|^2\notag\\
&~~~+\sum\limits_{j\in\mathcal{U}}\sum\limits_{k=1}^K\alpha_{i,k}(\hat{TN}^1_{e,j,k}+\hat{TN}^2_{e,j,k})+|c_{e,j}|^2+N_0),\\
&F_i^3=\log_2(\sum\limits_{j\neq i}\sum\limits_{k=1}^K\alpha_{i,k}(\hat{T}^1_{i,j,k}+\hat{T}^2_{i,j,k})
+|\hat{b}_{i,j}|^2\notag\\
&~~~+\sum\limits_{j\in\mathcal{U}}\sum\limits_{k=1}^K\alpha_{i,k}(\hat{TN}^1_{i,j,k}+\hat{TN}^2_{i,j,k})+|c_{i,j}|^2+N_0),\\
&F_i^4=\log_2(\sum\limits_{k=1}^K\alpha_{i,k}(\hat{T}^1_{e,i,k}+\hat{T}^2_{e,i,k})+|b_i|^2+\sum\limits_{j\neq i}\sum\limits_{k=1}^K\alpha_{i,k}(\hat{T}^1_{e,j,k}+\hat{T}^2_{e,j,k})\notag\\
&~~~+|\hat{b}_{e,j}|^2+\sum\limits_{j\in\mathcal{U}}\sum\limits_{k=1}^K\alpha_{i,k}(\hat{TN}^1_{e,j,k}+\hat{TN}^2_{e,j,k})+|c_{e,j}|^2+N_0).
\end{align}
}
In order to solve this subproblem, we first relax integer variable $\boldsymbol{\alpha}$, then we solve the problem by using convex optimization. After rounding the relaxed solution, we can get feasible $\boldsymbol{\alpha}$ for \textbf{Problem ~4a}. Similarly, we adopt the SCA method to construct global upper bounds of $F_i^3$ and $F_i^4$:
\begin{align}
F_i^3(\boldsymbol{\bar{\alpha}}_i)&\leq F^3_i(\boldsymbol{\bar{\alpha}}_i)+{\rm Tr}(\bigtriangledown_{\boldsymbol{\bar{\alpha}}_i}F^3_i(\boldsymbol{\bar{\alpha}}_i^{(t)})^H(\boldsymbol{\bar{\alpha}}_i-\boldsymbol{\bar{\alpha}}_i^{(t)}))\notag\\
&=\widetilde{F}_i^3(\boldsymbol{\bar{\alpha}}_i,\boldsymbol{\bar{\alpha}}_i^{(t)}),\\
F_i^4(\boldsymbol{\bar{\alpha}}_i)&\leq F^4_i(\boldsymbol{\bar{\alpha}}_i)+{\rm Tr}(\bigtriangledown_{\boldsymbol{\bar{\alpha}}_i}F^4_i(\boldsymbol{\bar{\alpha}}_i^{(t)})^H(\boldsymbol{\bar{\alpha}}_i-\boldsymbol{\bar{\alpha}}_i^{(t)}))\notag\\
&=\widetilde{F}_i^4(\boldsymbol{\bar{\alpha}}_i,\boldsymbol{\bar{\alpha}}_i^{(t)}),
\end{align}
where $\boldsymbol{\bar{\alpha}}_i=[\boldsymbol{\alpha}_{i,1}, ..., \boldsymbol{\alpha}_{i,K}]$. Thus, \textbf{Problem ~4a} can be transformed into a convex problem by introducing auxiliary variable $x$:
\begin{equation}
\textbf{Problem ~4b}:~~~\max\limits_{x,\boldsymbol{\alpha}}~~~ x
\end{equation}
\begin{align}
s.t.~~~
&0\leq x\leq F^1_i+F^2_i-\widetilde{F}^3_i-\widetilde{F}^4_i,\forall i\in\mathcal{U},\tag{C9}\\
&\sum\limits_k\alpha_{i,k}= 1, ~\alpha_{i,k}\in[0,1], ~\forall i\in\mathcal{U}. \label{C10}\tag{C10}
\end{align}
In this case, \textbf{Problem ~4b} becomes a general convex problem.

\section{Numerical Evaluation}
\begin{figure}[t]
  \centering
  \includegraphics[width=1.0\linewidth]{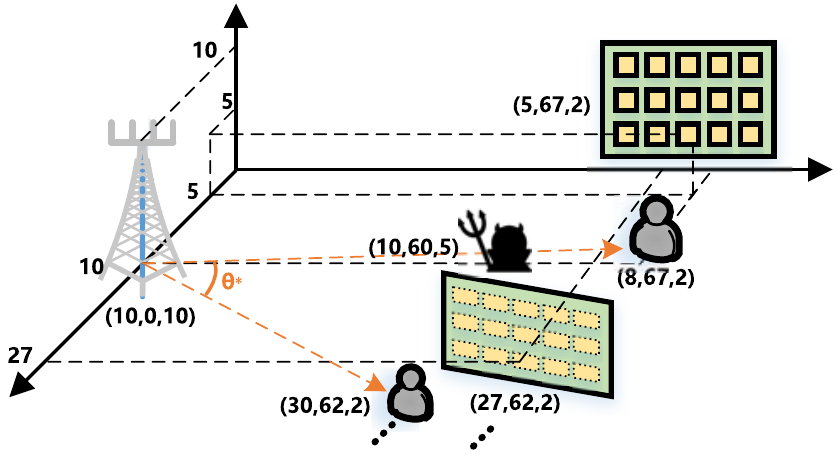}
  \caption{Overall setups for numerical evaluation.}
  \label{scenario}
\end{figure}

To evaluate the performance of the proposed scheme, we conduct a number of numerical evaluations in this section. The overall setup is shown in Fig. \ref{scenario}, we consider the base station is located at (10, 0, 10), IRSs and legitimate users are uniformly distributed around base station with a constant angle $\theta^*$. The first user and IRS are located at (5, 67, 5) and (8, 67, 2), respectively. Eve is located at (10, 60, 5) where in the middle between the base station and the first user.
We also assume that the direct channel between the base station and users are blocked by obstacles, which implies the channel state of the base station and a user is much worse than the channel state between the IRS and the user.
Specifically, the channels from base station to IRS/users/Eve are assumed to follow the distance-dependent path loss model, which can be generated by $\boldsymbol{h}=\sqrt{L_0d_{ab}^{-\beta}}\boldsymbol{h}^*$, where $d_{ab}$ denotes the distance from location $a$ to location $b$, and $\boldsymbol{h}^*$ is the small-scale fading component with Rician fading \cite{han2019intelligent,han2019large}:
\begin{equation}
\boldsymbol{h}^*=\sqrt{\frac{K}{K+1}}\boldsymbol{h}^*_{LoS}+\sqrt{\frac{1}{K+1}}\boldsymbol{h}^*_{NLoS},
\end{equation}
where $\boldsymbol{h}^*_{LoS}$ and $\boldsymbol{h}^*_{NLoS}$ represent the deterministic Line-of-Sight (LoS) and Rayleigh fading/Non-LoS (NLoS) components, respectively.
The LoS components are expressed as the responses of the $N$-elements uniform linear array $\boldsymbol{h}^*_{LoS}=\boldsymbol{a}_m(\theta)\boldsymbol{a}_n(\theta)^H$. The array response of an N-element IRS can be calculated by:

\begin{equation}
\boldsymbol{a}_m = \exp\left( j\frac{2\pi}{\lambda}d_{t}\left( m-1\right)\sin\phi_{LoS_1} \sin\theta_{LoS_1}\right), m=1,...,M,\notag
\end{equation}
\begin{equation}
\boldsymbol{a}_n = \exp(j\frac{2\pi}{\lambda}d_{r}(m-1)\sin\phi_{LoS_2} \sin\theta_{LoS_2}), n=1,...,N,\notag
\end{equation}
where $d_t$ and $d_r$ are the inter-antenna separation distance at the transmitter and receiver, $\phi_{LoS_1}$ and $\phi_{LoS_2}$ are the LoS azimuth at the base station and the IRS, and $\theta_{LoS_1}$ and $\theta_{LoS_2}$ are the angle of departure at the base station and the angle of arrival at the IRS, respectively.
The rest of parameter settings are listed in Table \ref{Simulation_Parameters}. Two baselines below are considered: 
\begin{itemize}
  \item \textit{Baseline 1}: Only beamforming is considered at the base station, and the IRS is not deployed in the system.
    \item \textit{Baseline 2}: Beamforming is considered at the base station, and only one IRS is deployed in the system.
\end{itemize}


\begin{table}[t]
  \centering
  \caption{Simulation Parameters}
  \label{Simulation_Parameters}
  \begin{tabular}{|c|p{4.8cm}|}
    \hline
     \textbf{Parameter} & \textbf{Value} \\\hline
     Carrier frequency & 2 GHz \\\hline
     \multirowcell{2}[-.2ex][c]{IRS configuration} & Uniform rectangular array with 5 elements in a row and N/5 columns with $3\lambda/8$ spacing\\\hline
     \multirowcell{2}[-.2ex][c]{Path loss exponent} & $\beta_{BU}=\beta_{BE}=5$, $\beta_{BI}=\beta_{IU}=\beta_{IE}=2$, respectively\\\hline
     \multirowcell{2}[-.2ex][c]{Rician channel factor} & $K_{BU}=K_{BE}=0$, $K_{BI}=K_{IU}=K_{IE}=\infty$, respectively\\\hline
     Path loss at 1 meter & $L_0=-30dB$\\\hline
     \multirowcell{2}[-.2ex][c]{Other parameters} & $N_0=-174$dBm, Tx = 4, $\delta=0.001$, $\theta^*=20^\circ$\\\hline
  \end{tabular}
\end{table}

\begin{figure}[t]
  \centering
  \includegraphics[width=1.0\linewidth]{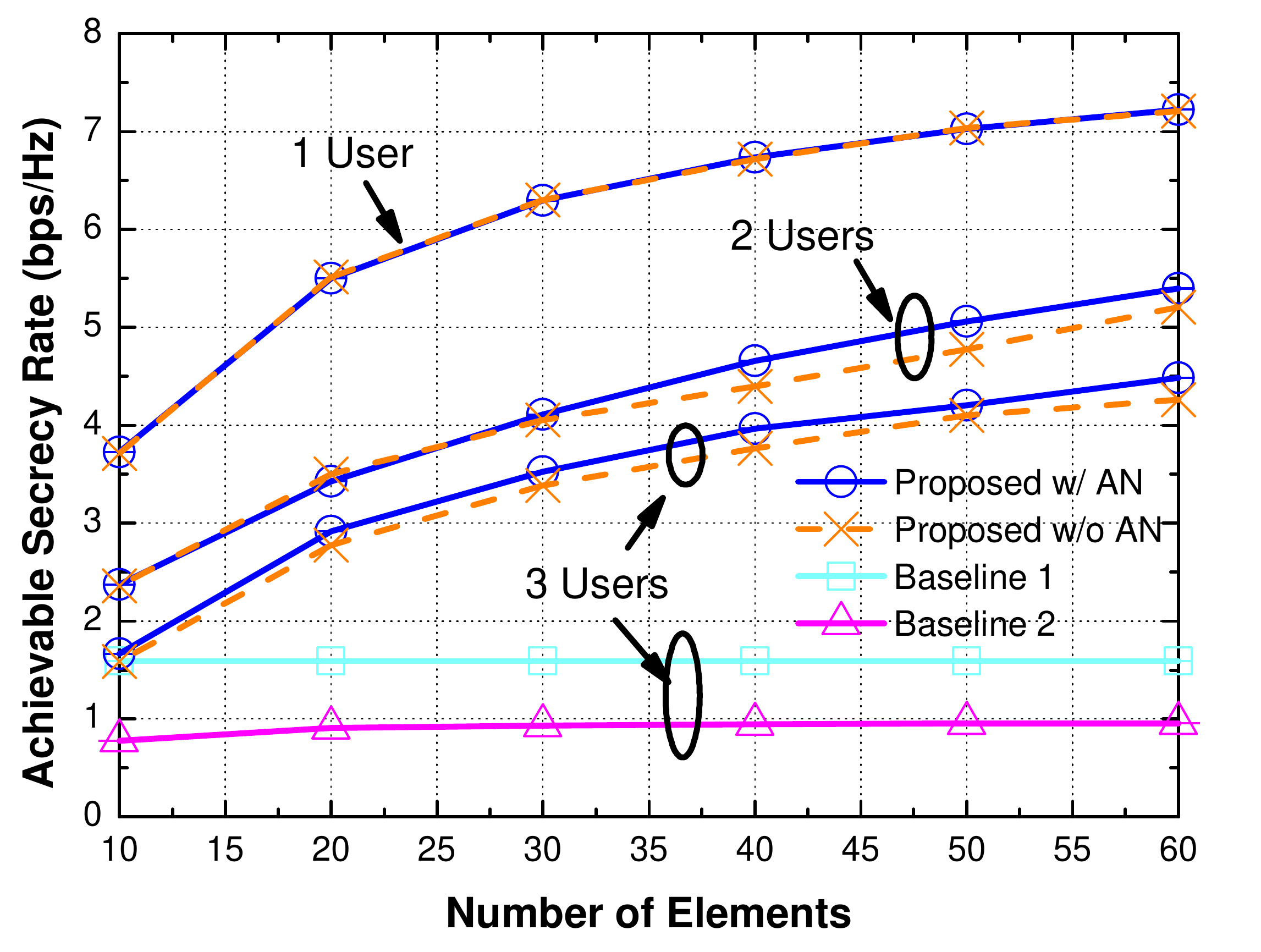}
  \caption{The achievable secrecy rate vs the number of elements ($P_{max}=40dBm$).}
  \label{ASRvsElements}
\end{figure}

Fig. \ref{ASRvsElements} shows the influence of different number of reflecting elements $N$ on each IRS. Due to the existence of obstacles, the LoS component is relatively poor for wireless transmissions between the base station and the user. When only one user is considered, the proposed scheme with AN has almost the same performance as the one without AN, which is is also verified in \cite{wu2019intelligent}. When there are 2 or more users, additional AN can help improve secrecy rate about 4-6\% especially with the increase in $N$.
Without the assistance of IRS and AN, baseline 1 has the worst performance compared with other schemes since the direct channel between the base station and the user is blocked. For fair comparison, we change $\beta_{BU}=\beta_{BE}=2$ for the beamforming scheme. The result also shows that the performance of beamforming scheme is relatively poor when there are multiple users.
For baseline 2, since users are distributed further apart from each other, only one IRS cannot satisfy the requirement for secure communications.

\begin{figure}[t]
  \centering
  \includegraphics[width=1.0\linewidth]{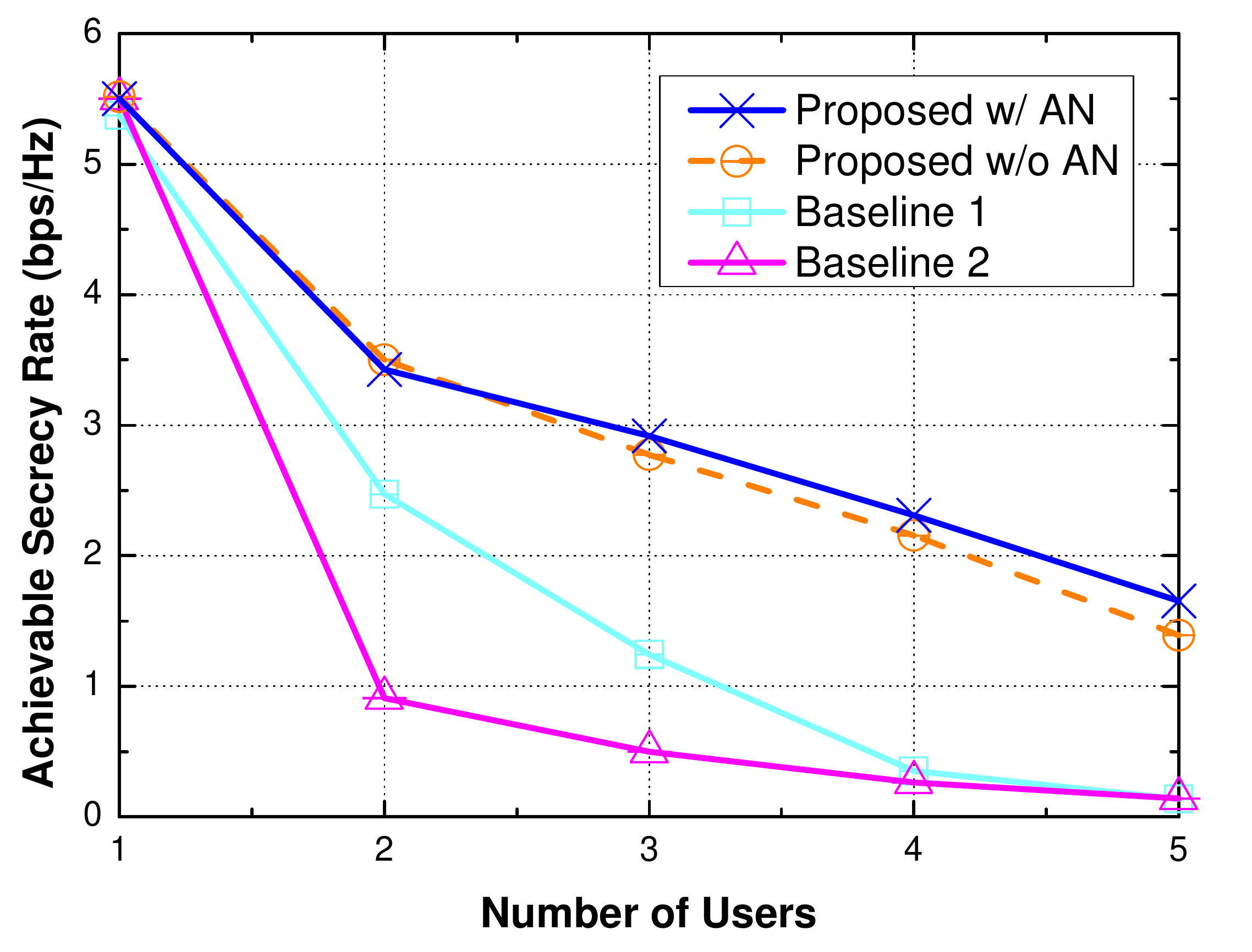}
  \caption{The achievable secrecy rate vs the number of users ($P_{max}=40dBm$ and $N=20$).}
  \label{ASRvsUsers}
\end{figure}

The achievable secrecy rate versus the number of users is shown in Fig. \ref{ASRvsUsers}. As we observe, the performance of all schemes in terms of achievable secrecy rate are degrading rapidly with the increase in the number of users. When there are more than 2 users, the proposed scheme perform better than AN-disabled scheme by up to 18.9\%. Here, for a fair comparison, we also set $\beta_{BU}=\beta_{BE}=2$ in baseline 1. The result also shows that the beamforming scheme in baseline 1 cannot deal with multiple users scenarios. Moreover, since the distance between the IRS and users significantly influences the performance of IRS-assisted schemes, we also set up a friendly scenario for baseline 2, i.e., all users are uniformly placed on the line from (8, 67, 2) to (8, 75, 2). When only a single IRS is deployed, the performance becomes even worse than that for baseline 1. The reason is that the environmental diversity provided by the IRS is very limited. If the overall performance is considered, e.g., the sum of secrecy rate, the system still can sacrifice a part of users' performance to achieve a better overall performance. If the worst performance in the system is considered as the objective, it becomes hard to optimize since each user matters. In this case, the algorithm tends to sacrifice the users who have the highest secrecy rate and make up for the users who have the worst secrecy rate, but the compensation is not significant enough due to the lack of environmental diversity. In this case, a poor performance is observed.

\begin{figure}[t]
  \centering
  \includegraphics[width=1.0\linewidth]{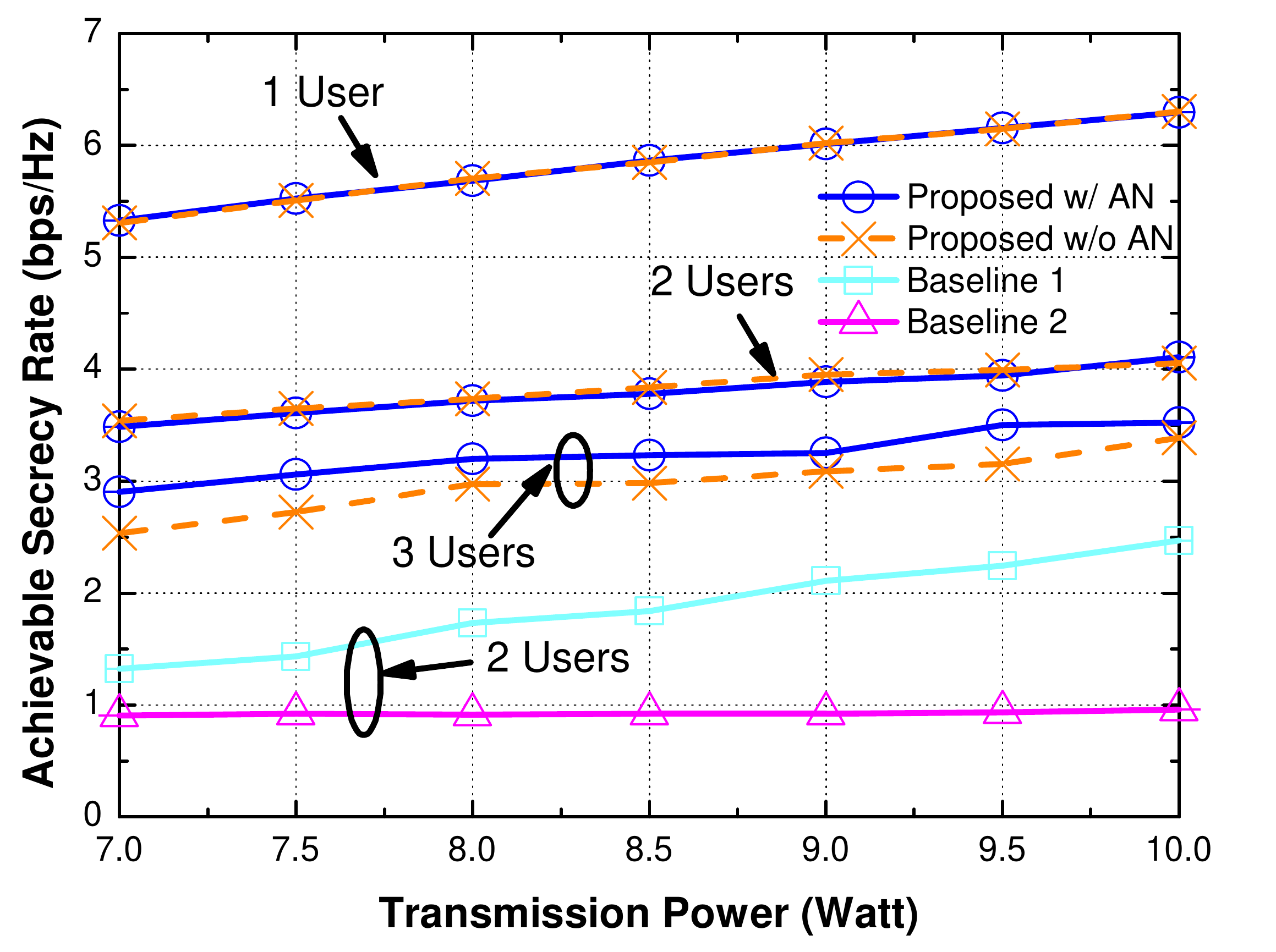}
  \caption{The achievable secrecy rate vs maximum transmission power ($N=30$).}
  \label{ASRvsPowerMax}
\end{figure}

The performance of achievable secrecy rate versus transmission power is shown in Fig. \ref{ASRvsPowerMax}. The maximum transmission power ranges from 7W (38.45dBm) to 10W (40dBm). With the increase in transmission power, the performance of all schemes increase linearly. Similar to the results in Fig. \ref{ASRvsUsers}, the proposed scheme outperforms the AN-disabled scheme when there are more than 2 users in the system. To have a fair comparison, we also consider LoS channel is not blocked by obstacle and set $\beta_{BU}=\beta_{BE}=2$ for baseline 1 with 2 users. However, the result shows that the performance of baseline 1 is much lower than IRS-assisted schemes. For baseline 2, since the performance is mainly limited by environmental diversity, it remains relatively steady and increases linearly from 0.9bps/Hz to 0.96bps/Hz with the increase in transmission power.

\begin{figure}[t]
  \centering
  \includegraphics[width=1.0\linewidth]{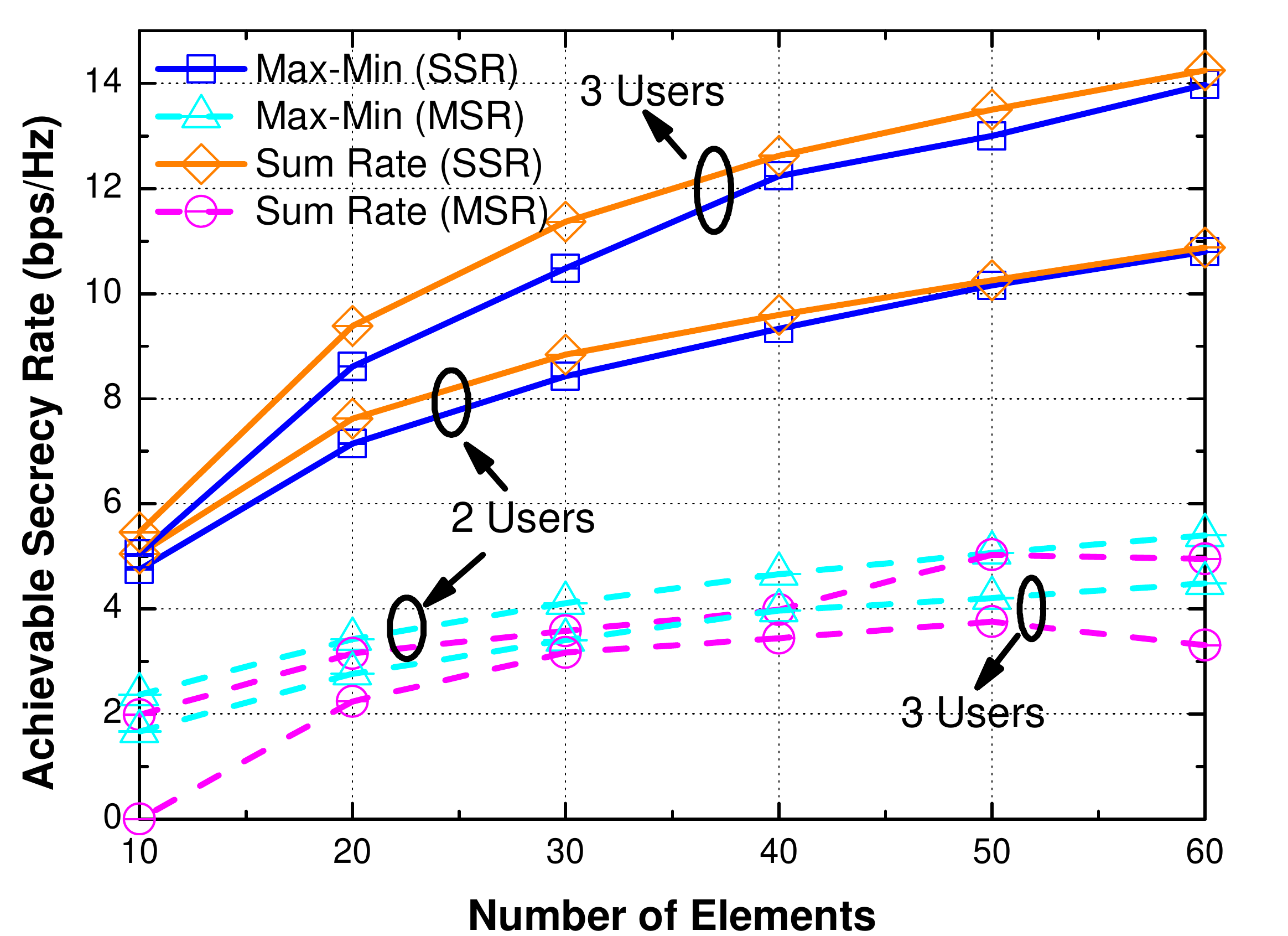}
  \caption{Max-min problem vs sum-rate problem in terms of the number of elements ($P_{max}=40dBm$).}
  \label{ASRvsSumRate}
\end{figure}

\begin{figure}[t]
  \centering
  \includegraphics[width=1.0\linewidth]{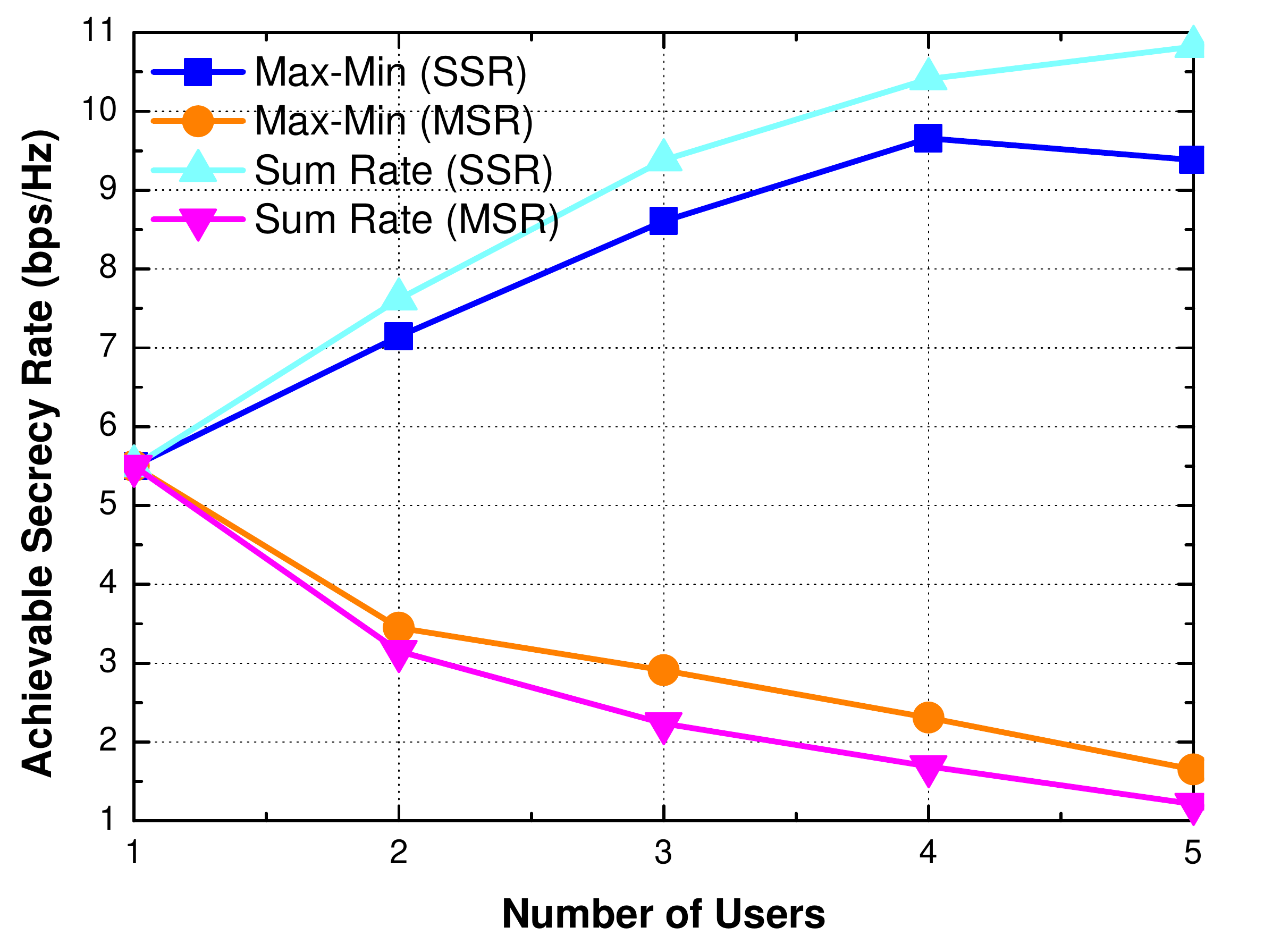}
  \caption{Max-min problem vs sum-rate problem in terms of the number of users ($P_{max}=40dBm, N=20$).}
  \label{ASRvsSumRate_NrofUsers}
\end{figure}

In general, overall security performance is a common objective considered in related works.
To compare the performance of the max-min problem proposed in this paper with the commonly studied sum-rate maximization, we plot Fig. \ref{ASRvsSumRate}-\ref{ASRvsSumRate_NrofUsers} to show the difference in terms of the minimum secrecy rate and the sum of secrecy rate, where the problem in (\ref{problem1}) with constraints (C1)-(C3) can be reformulated as:
\begin{equation}
\label{sumRate}
~~~\max\limits_{\boldsymbol{\bar{\omega}},\boldsymbol{\bar{z}},\boldsymbol{\bar{\Theta}},\boldsymbol{\bar{\alpha}}}
~\sum\limits_{i}[R_i^u-R_i^e]^+
\end{equation}
\begin{align}
s.t.~~~(C1)-(C3).\notag
\end{align}
Note that ``MSR'' and ``SSR'' in the legend represent the minimum secrecy rate and the sum of secrecy rate, respectively.
As shown in Fig. \ref{ASRvsSumRate}, for the performance in terms of the minimum secrecy rate, the gap between two objectives can vary rapidly with different number of elements, especially for the scenarios with more users, which implies that sum-rate objective can hardly guarantee the worst secrecy rates for different users.
For the performance in terms of the sum of secrecy rate,
with the increase in the number of elements, the overall sum of secrecy rates of these two objective tends to converge and have similar performance. This phenomenon indicates that a max-min problem can achieve better minimum secrecy rate and also reach similar performance in overall secrecy rate for a large-scale IRS-assisted system.
Meanwhile, in Fig. \ref{ASRvsSumRate_NrofUsers}, the sum of secrecy rate increases with the number of users. Even though it can sacrifice some users' performance to improve overall performance, the curve shows that the gain becomes less and the sum secrecy rate reaches a threshold with the increase in the number of users, which represents the maximum secrecy capacity in the system. For the gap between two different objectives, it also becomes larger with the increase in the number of users, which is reasonable since the solution space becomes larger with more users in the system, and different solutions obtained from the aforementioned objectives do impact more users.
\section{Conclusion}

In this paper, we have focused on physical layer security in wireless systems with IRSs, and investigated a max-min problem regarding secrecy rate under one typical eavesdropper scenario. By placing multiple collaborative IRSs in complex environment, the base station could leverage the environmental diversity to achieve significant improvement in terms of secrecy rate through joint optimization of beamforming and phase shift on the IRS. Based on our numerical evaluation, when multiple users are considered, the additional AN has been proven to effectively create interference at the eavesdropper and further improve the performance in terms of secrecy rate.
Compared with sum-rate maximization studied in related works, we have found out that the performance of the proposed max-min problem converges to that of sum-rate problem in terms of the sum of secrecy rate when increasing the number of elements on IRSs.
In the future, we plan to extend our study by considering a general adversary model and explore the specific collaborative protocols/mechanism among multiple IRSs.


%

%

%
%

\ifCLASSOPTIONcaptionsoff
  \newpage
\fi



%
\bibliographystyle{IEEEtran}
\footnotesize
\bibliography{bibliography-IRS}

\begin{thebibliography}{10}
\providecommand{\url}[1]{#1}
\csname url@samestyle\endcsname
\providecommand{\newblock}{\relax}
\providecommand{\bibinfo}[2]{#2}
\providecommand{\BIBentrySTDinterwordspacing}{\spaceskip=0pt\relax}
\providecommand{\BIBentryALTinterwordstretchfactor}{4}
\providecommand{\BIBentryALTinterwordspacing}{\spaceskip=\fontdimen2\font plus
\BIBentryALTinterwordstretchfactor\fontdimen3\font minus
  \fontdimen4\font\relax}
\providecommand{\BIBforeignlanguage}[2]{{%
\expandafter\ifx\csname l@#1\endcsname\relax
\typeout{** WARNING: IEEEtran.bst: No hyphenation pattern has been}%
\typeout{** loaded for the language `#1'. Using the pattern for}%
\typeout{** the default language instead.}%
\else
\language=\csname l@#1\endcsname
\fi
#2}}
\providecommand{\BIBdecl}{\relax}
\BIBdecl

\bibitem{chraiti2017achieving}
M.~Chraiti, A.~Ghrayeb, and C.~Assi, ``Achieving full secure degrees-of-freedom
  for the {MISO} wiretap channel with an unknown eavesdropper,'' \emph{IEEE
  Transactions on Wireless Communications}, vol.~16, no.~11, pp. 7066--7079,
  2017.

\bibitem{li2013transmit}
Q.~Li, M.~Hong, H.-T. Wai, Y.-F. Liu, W.-K. Ma, and Z.-Q. Luo, ``Transmit
  solutions for {MIMO} wiretap channels using alternating optimization,''
  \emph{IEEE Journal on Selected Areas in Communications}, vol.~31, no.~9, pp.
  1714--1727, 2013.

\bibitem{he2014mimo}
X.~He and A.~Yener, ``{MIMO} wiretap channels with unknown and varying
  eavesdropper channel states,'' \emph{IEEE Transactions on Information
  Theory}, vol.~60, no.~11, pp. 6844--6869, 2014.

\bibitem{wang2019secure}
H.-M. Wang, X.~Zhang, Q.~Yang, and T.~A. Tsiftsis, ``Secure users oriented
  downlink {MISO NOMA},'' \emph{IEEE Journal of Selected Topics in Signal
  Processing}, vol.~13, no.~3, pp. 671--684, 2019.

\bibitem{guan2020intelligent}
X.~Guan, Q.~Wu, and R.~Zhang, ``Intelligent reflecting surface assisted secrecy
  communication: Is artificial noise helpful or not?'' \emph{IEEE Wireless
  Communications Letters}, 2020.

\bibitem{zou2015improving}
Y.~Zou, J.~Zhu, X.~Wang, and V.~C. Leung, ``Improving physical-layer security
  in wireless communications using diversity techniques,'' \emph{IEEE Network},
  vol.~29, no.~1, pp. 42--48, 2015.

\bibitem{shiu2011physical}
Y.-S. Shiu, S.~Y. Chang, H.-C. Wu, S.~C.-H. Huang, and H.-H. Chen, ``Physical
  layer security in wireless networks: A tutorial,'' \emph{IEEE wireless
  Communications}, vol.~18, no.~2, pp. 66--74, 2011.

\bibitem{zhu2014joint}
F.~Zhu, F.~Gao, M.~Yao, and H.~Zou, ``Joint information-and jamming-beamforming
  for physical layer security with full duplex base station,'' \emph{IEEE
  Transactions on Signal Processing}, vol.~62, no.~24, pp. 6391--6401, 2014.

\bibitem{zhu2015improving}
F.~Zhu and M.~Yao, ``Improving physical-layer security for {CRN}s using
  {SINR}-based cooperative beamforming,'' \emph{IEEE Transactions on Vehicular
  Technology}, vol.~65, no.~3, pp. 1835--1841, 2015.

\bibitem{di2020smart}
M.~Di~Renzo, A.~Zappone, M.~Debbah, M.-S. Alouini, C.~Yuen, J.~de~Rosny, and
  S.~Tretyakov, ``Smart radio environments empowered by reconfigurable
  intelligent surfaces: How it works, state of research, and road ahead,''
  \emph{arXiv preprint arXiv:2004.09352}, 2020.

\bibitem{di2019smart}
M.~Di~Renzo, M.~Debbah, D.-T. Phan-Huy, A.~Zappone, M.-S. Alouini, C.~Yuen,
  V.~Sciancalepore, G.~C. Alexandropoulos, J.~Hoydis, H.~Gacanin \emph{et~al.},
  ``Smart radio environments empowered by reconfigurable ai meta-surfaces: An
  idea whose time has come,'' \emph{EURASIP Journal on Wireless Communications
  and Networking}, vol. 2019, no.~1, pp. 1--20, 2019.

\bibitem{wu2019towards}
Q.~Wu and R.~Zhang, ``Towards smart and reconfigurable environment: Intelligent
  reflecting surface aided wireless network,'' \emph{IEEE Communications
  Magazine}, 2019.

\bibitem{wu2019intelligent}
------, ``Intelligent reflecting surface enhanced wireless network via joint
  active and passive beamforming,'' \emph{IEEE Transactions on Wireless
  Communications}, vol.~18, no.~11, pp. 5394--5409, 2019.

\bibitem{zhang2020augmenting}
L.~Zhang, L.~Yan, B.~Lin, H.~Ding, Y.~Fang, and X.~Fang, ``Augmenting
  transmission environments for better communications: tunable reflector
  assisted mmwave wlans,'' \emph{IEEE Transactions on Vehicular Technology},
  vol.~69, no.~7, pp. 7416--7428, 2020.

\bibitem{zhang2019tunable}
L.~Zhang, L.~Yan, B.~Lin, Y.~Fang, and X.~Fang, ``Tunable reflectors enabled
  environment augmentation for better mmwave wlans,'' in \emph{2019 IEEE/CIC
  International Conference on Communications in China (ICCC)}.\hskip 1em plus
  0.5em minus 0.4em\relax IEEE, 2019, pp. 7--12.

\bibitem{cui2019secure}
M.~Cui, G.~Zhang, and R.~Zhang, ``Secure wireless communication via intelligent
  reflecting surface,'' \emph{IEEE Wireless Communications Letters}, vol.~8,
  no.~5, pp. 1410--1414, 2019.

\bibitem{yu2019enabling}
X.~Yu, D.~Xu, and R.~Schober, ``Enabling secure wireless communications via
  intelligent reflecting surfaces,'' \emph{arXiv preprint arXiv:1904.09573},
  2019.

\bibitem{dong2020secure}
L.~Dong and H.-M. Wang, ``Secure {MIMO} transmission via intelligent reflecting
  surface,'' \emph{IEEE Wireless Communications Letters}, 2020.

\bibitem{lyu2020irs}
B.~Lyu, D.~T. Hoang, S.~Gong, D.~Niyato, and D.~I. Kim, ``{IRS}-based wireless
  jamming attacks: When jammers can attack without power,'' \emph{arXiv
  preprint arXiv:2001.01887}, 2020.

\bibitem{xu2019resource}
D.~Xu, X.~Yu, Y.~Sun, D.~W.~K. Ng, and R.~Schober, ``Resource allocation for
  secure irs-assisted multiuser {MISO} systems,'' in \emph{2019 IEEE Globecom
  Workshops (GC Wkshps)}.\hskip 1em plus 0.5em minus 0.4em\relax IEEE, 2019,
  pp. 1--6.

\bibitem{zheng2019intelligent}
B.~Zheng and R.~Zhang, ``Intelligent reflecting surface-enhanced ofdm: Channel
  estimation and reflection optimization,'' \emph{IEEE Wireless Communications
  Letters}, vol.~9, no.~4, pp. 518--522, 2019.

\bibitem{li2020energy}
J.~Li, K.~Xue, D.~S. Wei, J.~Liu, and Y.~Zhang, ``Energy efficiency and traffic
  offloading optimization in integrated satellite/terrestrial radio access
  networks,'' \emph{IEEE Transactions on Wireless Communications}, vol.~19,
  no.~4, pp. 2367--2381, 2020.

\bibitem{razaviyayn2014successive}
M.~Razaviyayn, ``Successive convex approximation: Analysis and applications,''
  2014.

\bibitem{sun2017optimal}
Y.~Sun, D.~W.~K. Ng, Z.~Ding, and R.~Schober, ``Optimal joint power and
  subcarrier allocation for full-duplex multicarrier non-orthogonal multiple
  access systems,'' \emph{IEEE Transactions on Communications}, vol.~65, no.~3,
  pp. 1077--1091, 2017.

\bibitem{alvarado2014new}
A.~Alvarado, G.~Scutari, and J.-S. Pang, ``A new decomposition method for
  multiuser {DC}-programming and its applications,'' \emph{IEEE Transactions on
  Signal Processing}, vol.~62, no.~11, pp. 2984--2998, 2014.

\bibitem{simchowitz2018course}
M.~Simchowitz, ``Course notes for ee227c (spring 2018): Convex optimization and
  approximation,'' 2018.

\bibitem{luo2010semidefinite}
Z.-Q. Luo, W.-K. Ma, A.~M.-C. So, Y.~Ye, and S.~Zhang, ``Semidefinite
  relaxation of quadratic optimization problems,'' \emph{IEEE Signal Processing
  Magazine}, vol.~27, no.~3, pp. 20--34, 2010.

\bibitem{ma2010semidefinite}
W.-K.~K. Ma, ``Semidefinite relaxation and its applications in signal
  processing and communications,'' \emph{IEEE SIGNAL PROCESSING MAGAZINE}, vol.
  1053, no. 5888/10, 2010.

\bibitem{ma2012semidefinite}
W.-K. Ma, ``Semidefinite relaxation and its applications in signal processing
  and communications,'' \emph{MIIS Tutorial}, July 2012.

\bibitem{sedumi}
``{S}e{D}u{M}i {S}oftware,'' \url{http://sedumi.ie.lehigh.edu/}, accessed
  September 1, 2020.

\bibitem{grant2014cvx}
M.~Grant and S.~Boyd, ``{CVX}: Matlab software for disciplined convex
  programming, version 2.1,'' 2014.

\bibitem{nasir2015joint}
A.~A. Nasir, D.~T. Ngo, X.~Zhou, R.~A. Kennedy, and S.~Durrani, ``Joint
  resource optimization for multicell networks with wireless energy harvesting
  relays,'' \emph{IEEE Transactions on Vehicular Technology}, vol.~65, no.~8,
  pp. 6168--6183, 2015.

\bibitem{wang2012successive}
T.~Wang and L.~Vandendorpe, ``Successive convex approximation based methods for
  dynamic spectrum management,'' in \emph{2012 IEEE International Conference on
  Communications (ICC)}.\hskip 1em plus 0.5em minus 0.4em\relax IEEE, 2012, pp.
  4061--4065.

\bibitem{han2019intelligent}
H.~Han, J.~Zhao, D.~Niyato, M.~Di~Renzo, and Q.-V. Pham, ``Intelligent
  reflecting surface aided network: Power control for physical-layer
  broadcasting,'' \emph{arXiv preprint arXiv:1910.14383}, 2019.

\bibitem{han2019large}
Y.~Han, W.~Tang, S.~Jin, C.-K. Wen, and X.~Ma, ``Large intelligent
  surface-assisted wireless communication exploiting statistical {CSI},''
  \emph{IEEE Transactions on Vehicular Technology}, vol.~68, no.~8, pp.
  8238--8242, 2019.

\end{thebibliography}

%




\end{document}